\documentclass[final,5p,twocolumn]{elsarticle}

\usepackage[utf8]{inputenc}
\usepackage[T1]{fontenc}
\usepackage{bm}
\usepackage{amssymb}
\usepackage{gensymb}
\usepackage{amsmath}
\usepackage{amsthm}
\usepackage{textcomp}
\usepackage{hyperref}
\usepackage{afterpage}
\usepackage{comment}
\usepackage{graphicx} 
\usepackage{subfigure}
\usepackage{subcaption}
\usepackage{float} 
\usepackage{url}
\usepackage{xcolor} 
\usepackage{enumitem} 
\usepackage{adjustbox}
\usepackage{titlesec}
\usepackage{stfloats}
\usepackage{setspace}
\usepackage{multirow} 
\usepackage{siunitx} 
\usepackage[version=4]{mhchem} 
\usepackage[capitalise, nameinlink, noabbrev]{cleveref}
\usepackage{soul}
\usepackage{ragged2e}
\usepackage[most]{tcolorbox}
\usepackage{multicol}
\usepackage{tabularx}
\usepackage{booktabs}
\usepackage{longtable}
\usepackage{threeparttable}
\usepackage{makecell}
\newcolumntype{L}[1]{>{\raggedright\arraybackslash}m{#1}}
\newcolumntype{C}[1]{>{\centering\arraybackslash}m{#1}}
\usepackage{booktabs}
\usepackage{bm}
\hypersetup{
    colorlinks=true,
    linkcolor=blue,
    filecolor=magenta,      
    urlcolor=blue,
    } 
\usepackage{etoolbox}

\usepackage{lineno}

\begin{document}
\newpage
\begin{frontmatter}
\title{\textbf{\LARGE Vacancy Engineering in Metals and Alloys}}

\author[inst1,inst2]{Sreenivas Raguraman\corref{cor1}}
\ead{sragura1@jhu.edu}
\author[inst3]{Homero Reyes Pulido}
\author[inst4]{Christopher Hutchinson}
\author[inst5]{Arun Devaraj}
\author[inst6]{Marc H. Weber}
\author[inst1,inst2,inst3,inst7]{Michael L. Falk}
\author[inst1,inst2,inst7]{Timothy P. Weihs\corref{cor1}}
\ead{weihs@jhu.edu}

\affiliation[inst1]{organization={Department of Materials Science and Engineering, Johns Hopkins University},state={MD}, country={U.S.A}}
\affiliation[inst2]{organization={Hopkins Extreme Materials Institute, Johns Hopkins University},state={MD},country={U.S.A}}
\cortext[cor1]{Corresponding authors}
\affiliation[inst3]{organization={Department of Physics and Astronomy, Johns Hopkins University},state={MD},country={U.S.A}}
\affiliation[inst4]{organization={Department of Materials Science and Engineering, Monash University},state={VIC},country={Australia}}
\affiliation[inst5]{organization={Physical and Computational Sciences Directorate, Pacific Northwest National University},state={WA},country={U.S.A}}
\affiliation[inst6]{organization={Institute of Materials Research, Washington State University, Pullman}, state={WA},country={U.S.A}}
\affiliation[inst7]{organization={Department of Mechanical Engineering, Johns Hopkins University}, state={MD},country={U.S.A}}
\begin{abstract}
Vacancy engineering, the intentional control of atomic-scale vacancies in metals and alloys, is emerging as a powerful yet underexplored strategy for tailoring microstructures and optimizing performance across diverse applications. By enabling excess vacancy populations through quenching, severe deformation, thermomechanical treatments, or additive manufacturing, new microstructures can be obtained that achieve unique combinations of strength, ductility, fatigue life, corrosion resistance, and conductivity. Vacancies are distinct among lattice defects: they are non-conserved entities essential for solute diffusion, yet variably coupled to solutes, dislocations, and phase boundaries. They can accelerate transformations such as nucleation and precipitation or retard kinetics when trapped in clusters, and their transient trapping and release can drive microstructural evolution across time and length scales. This Review synthesizes recent advances in generating, modeling, and characterizing vacancies, highlighting their role in diffusion, precipitation, and phase stability. Case studies in lightweight, high-temperature, fatigue-resistant, electrical, and biomedical materials demonstrate the broad potential of vacancy control. We conclude by emphasizing the opportunity for the metallurgical community to fully exploit excess vacancies as controllable, design-relevant defects that enable new pathways for microstructure and property optimization in next-generation alloys.
\end{abstract}

\end{frontmatter}

\section*{Introduction}\label{sec:intro}
Vacancy engineering, the purposeful creation and manipulation of atomic vacancies to control material properties, has become an increasingly powerful concept across different classes of materials. In functional systems such as thermoelectrics, semiconductors, electrocatalysts, and bimetallic oxides, engineered vacancy populations are used to tune carrier concentration, electronic structure, and chemical reactivity, thereby enhancing performance and efficiency \cite{zhang2021advancing,wang2021vacancy,wu2021recent,mao2024situ}. These advances have established vacancies as dynamic design variables that can couple structure, chemistry, and functionality. In contrast, the deliberate manipulation of vacancies in structural metals and alloys, where vacancies intrinsically govern diffusion, phase transformations, and defect interactions, remains comparatively underexplored. Vacancies have long been recognized as essential lattice defects that control substitutional diffusion and phase transformations in metals, yet recent experimental and computational studies reveal their broader capacity to influence microstructural evolution through solute clustering, phase stability, and defect interactions. Harnessing these effects through the purposeful generation and control of non-equilibrium vacancy populations represents a largely untapped route to designing defect-mediated processes that can tailor alloy microstructures and achieve new levels of performance.

Traditional alloy design has primarily relied on manipulating solute distributions, dislocation densities, grain and phase boundaries, and secondary phases to tailor properties. Direct manipulation of vacancies, despite their pivotal role in governing precipitation, solute redistribution, creep, and transformation pathways, has not yet been systematically integrated into metallurgical design strategies. This gap becomes particularly significant in systems with sluggish precipitation kinetics, such as magnesium alloys \cite{ma2019dynamic}, and in multifunctional materials that require both mechanical robustness and controlled degradation, such as biodegradable implants \cite{raguraman_simultaneous_2025} and high-temperature components \cite{oleksak2018role,luo2025determinants}. Even more complex behavior is expected in high-entropy and compositionally complex alloys, where lattice distortion, chemical short-range order, and local energy fluctuations modify vacancy formation energies and diffusion pathways \cite{roy2022vacancy,luo2025determinants}. Although the fundamental role of vacancies in diffusion and defect-mediated transport is well established, translating this understanding into predictive control over microstructure remains a central challenge. Bridging this divide requires deciphering how alloy chemistry, processing history, and vacancy kinetics collectively dictate microstructural evolution, paving the way for deliberate, vacancy-informed alloy design.

Similar to grain boundary engineering, which focuses on modifying boundary character rather than removing boundaries, vacancy engineering involves the deliberate control of vacancy population, distribution, and stability to guide microstructural evolution. Vacancies are distinct from other defects: they are non-conserved entities essential for solute diffusion yet dynamically coupled to solutes, dislocations, and phase boundaries. Depending on their state, they can accelerate transformations such as precipitation and clustering or slow them when trapped in complexes. They also enable dislocation climb and boundary motion and can be temporarily immobilized and later released to drive structural change. Here, we define vacancy engineering as the purposeful manipulation of these dynamic behaviors, which establishes a coherent framework for exploiting vacancies as active design variables in metals and alloys.

Foundational work in the mid-20th century, especially by Simmons and Balluffi \cite{simmons1960measurement,simmons1962measurement}, established vacancies as measurable thermodynamic entities by relating equilibrium concentrations to lattice expansion through X-ray diffraction and dilatometry. These efforts reframed vacancies as experimentally accessible, temperature-dependent features. Building on this legacy, techniques such as positron annihilation spectroscopy (PAS) \cite{selim2021positron,weber2014vacancies}, atom probe tomography (APT) \cite{chen2025using,wang2024effect}, 3D-field ion microscopy \cite{katnagallu2018advanced,dagan2017automated}, and four-dimensional scanning transmission electron microscopy (4D-STEM) \cite{mills2023nanoscale,yang2023one} now probe vacancy-related phenomena. Complementary computational tools, including first-principles calculations \cite{wolverton_solutevacancy_2007,shin2010first}, molecular dynamics (MD) \cite{xu2003molecular,huang1989pipe}, and Kinetic Monte Carlo (KMC) simulations \cite{battaile2008kinetic,voter2007introduction} model vacancy formation, migration, and interactions. The growing convergence between experiment and simulation is enabling more accurate, mechanistic insights into microstructure evolution, positioning the field to shift from descriptive studies to proactive microstructure design informed by vacancy engineering.

This Review outlines the evolving landscape of vacancy engineering as a strategy for controlling microstructure and properties in metals and alloys. We begin by examining how excess vacancies can be generated and stabilized, and how their thermodynamic and kinetic behavior governs clustering, solute interactions, and phase stability. We then synthesize recent advances in characterization and modeling, with an emphasis on integrating multimodal experimental workflows and atomistic simulations. Building on these foundations, we present application-focused case studies that illustrate how vacancy control enhances performance in lightweight, high-temperature, fatigue-resistant, electrical, and biomedical materials. Finally, we identify key experimental, conceptual, and integration challenges that must be addressed to realize a predictive, design-oriented approach to vacancy engineering across length scales and material classes. By unifying previous studies under a coherent framework, we aim to promote vacancy engineering as a tool for designing next-generation structural and functional alloys.

\section*{Mechanisms of generating non-equilibrium concentration of vacancies}

\begin{figure*}[h!]
    \centering
    \includegraphics[width=1\linewidth]{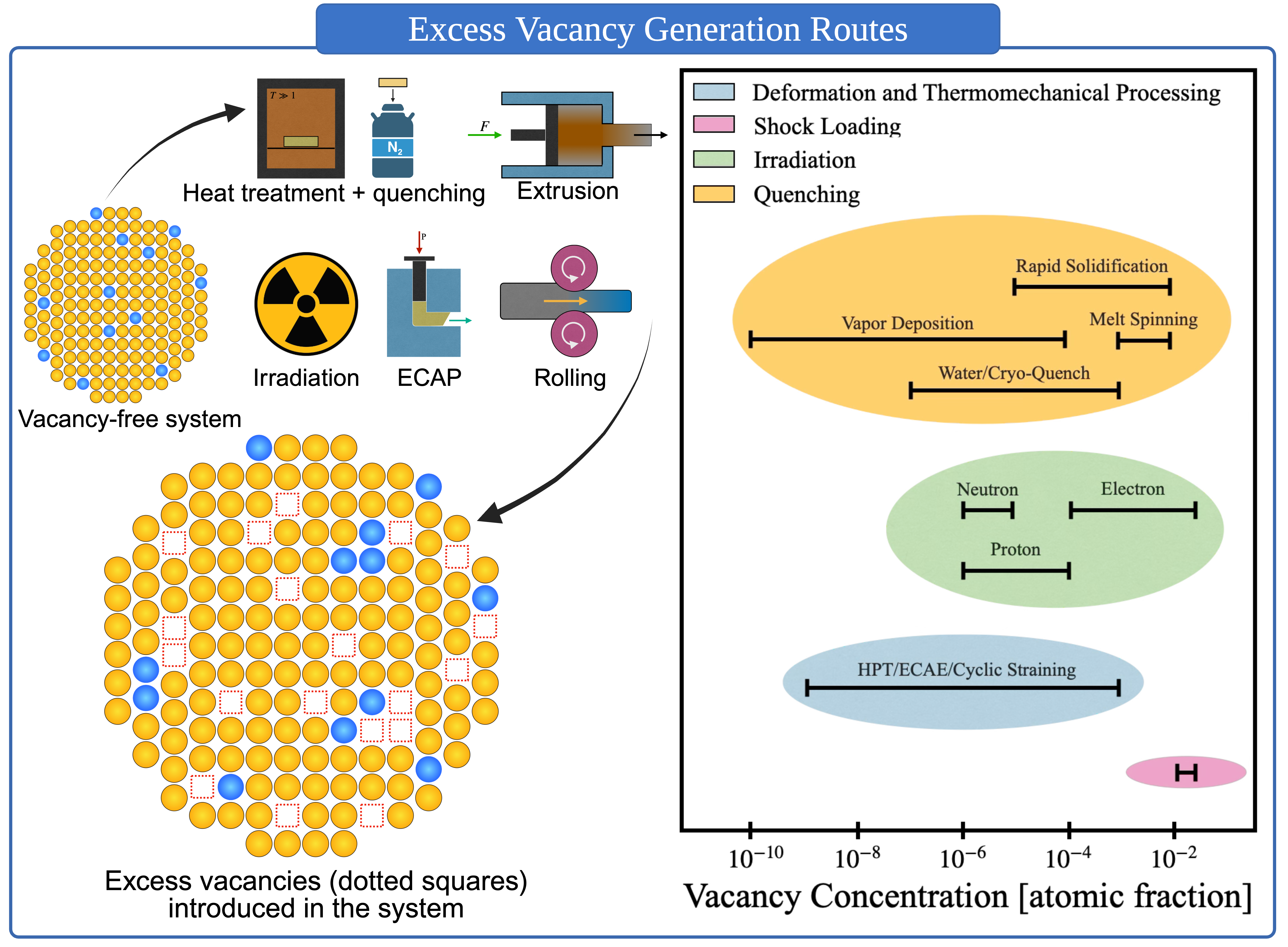}
    \caption{\justifying \textbf{Processing routes and mechanisms for generating excess vacancies in metals and alloys.} The schematic on the left illustrates how different processing routes introduce excess vacancies into metallic lattices through distinct physical mechanisms. The plot on the right compares reported non-equilibrium vacancy concentrations, showing orders-of-magnitude variation among routes. Irradiation, shock loading, and rapid solidification generally yield the highest vacancy fractions \cite{boleininger2023microstructure,pedchenko1969concerning,majeed2025vacancy,hillert2002trapping,haraguchi2003determination,yang2010anomaly,gavsparova2024positron}, while severe plastic deformation (SPD) and solid state quenching (e.g. water and liquid nitrogen) produce intermediate levels \cite{ma2019dynamic,cizek2019development,sun2019precipitation,wu2022freezing,chen2023investigation}. Vapor deposition-based processes, span a broad range depending on growth kinetics and substrate conditions \cite{zhou1997vacancy}. Together, these pathways demonstrate how excess vacancies can be engineered to control diffusion, phase stability, precipitation behavior, and mechanical performance.}
    \label{fig:Vacancy_Generation}
    \vspace{-1em}
\end{figure*}

Vacancy concentrations in metals can exceed equilibrium levels when lattice atoms are displaced or diffusion is suppressed, particularly during rapid thermal, mechanical, or irradiation-driven processing. \textbf{Figure~\ref{fig:Vacancy_Generation}} categorizes these vacancy-generating routes into deformation-, thermal-, and energetic-driven mechanisms, and compares their relative vacancy magnitudes.

While each method induces supersaturation through different means, the spatial distribution and persistence of vacancies vary considerably. This section first outlines irradiation- and shock-induced mechanisms for context, then focuses on the thermomechanical and thermal routes most relevant to structural alloy design.

\subsection*{Irradiation and Shock Loading}
In high-energy environments such as nuclear reactors and space systems, vacancy formation is dominated by displacement cascades initiated by energetic particles \cite{kapinos1995model}. These cascades eject atoms from lattice sites, generating Frenkel pairs comprising vacancies and self-interstitials. When irradiation is sustained, vacancies coalesce into clusters or voids, contributing to swelling, embrittlement, and microstructural instability \cite{dubinko2011radiation,wurschum1996characterization,tanaka2019irradiation}. For example, neutron irradiation can lead to void nucleation, compromising ductility and dimensional control \cite{laidler1972nucleation,yang2010anomaly,zinkle2014designing}.

Similarly, shock waves and high-strain-rate deformation produce non-equilibrium vacancies through dislocation interactions. Localized heating and dislocation glide create jogs that pin segments and induce vacancy generation \cite{reina2011nanovoid,jiang2021effects,ho2007energetics}. In extreme scenarios, such as ballistic impact, shock-induced supersaturation of vacancies can reach sufficiently high levels to trigger local transient melting or amorphization, followed by rapid quenching. The resulting vacancies diffuse toward grain boundaries, where they may annihilate or coalesce into nanovoids, thereby facilitating intergranular fracture and spallation \cite{adibi2020evolving,xu2016ballistic,holian1998plasticity}. In nickel-based alloys, these damage mechanisms are linked to void nucleation and dynamic failure under stress wave propagation \cite{fukai2001superabundant,smalinskas1993study}.

Irradiation and shock loading are powerful methods for generating excess vacancies, providing fundamental insights into defect formation and evolution under extreme conditions. However, the vacancy populations produced by such routes are often transient and spatially non-uniform, reflecting localized energy deposition and rapid relaxation. This Review, therefore, focuses on practical, design-relevant processing methods, including quenching and deformation, that enable more isotropic and controllable vacancy generation and stabilization, offering greater potential for integration into alloy design and scalable manufacturing.

\subsection*{Quenching from High Temperatures}
Rapid cooling from elevated temperatures is one of the most established routes for generating non-equilibrium vacancies in metals \cite{adams1993void,kino1967vacancies}. In the solid state, vacancies equilibrate with the high-temperature lattice, but rapid quenching suppresses their diffusion to sinks such as dislocations, grain boundaries, and surfaces, leading to supersaturation \cite{lukavc2013vacancy,kimura1959quenched,morris1998quenching}. These retained vacancies profoundly influence subsequent microstructural evolution by accelerating solute clustering and precipitation. For instance, in aluminum alloys, quenching from the solution-treated state traps vacancy concentrations several orders of magnitude above equilibrium, enabling rapid solute migration and early-stage cluster formation that enhance age-hardening kinetics \cite{silcock1959effect,khellaf2002quenching,chen2023investigation}. Conversely, slower cooling promotes vacancy annihilation at grain boundaries, forming precipitate-free zones (PFZs) that degrade ductility and fatigue resistance \cite{hirsch1958dislocation,zhang2020training}.

Liquid-phase cooling, including rapid solidification such as in additive manufacturing (AM) and melt spinning, involves even higher cooling rates ($10^3$--$10^6$ K/s), enabling significant vacancy retention in powders and as-built microstructures \cite{haraguchi2005vacancy,zhou2019heat,shakil2021additive}. In AM-processed AlSi10Mg alloys, such vacancies affect solute partitioning, residual stress evolution, and precipitation during post-processing \cite{shakil2021additive}. Yet the role of vacancies in AM systems remains understudied, with current efforts focused largely on porosity and microstructure refinement. Greater attention to vacancy behavior could unlock new strategies for tailoring microstructures via thermal treatments and aging protocols.

Vacancy supersaturation can also arise during vapor-phase processing methods such as physical or chemical vapor deposition (PVD, CVD). These techniques operate under low adatom mobility, which limits defect annihilation and promotes vacancy-rich growth regimes \cite{gruber2011strain,lorenzin2022stress,abadias2018stress}. In Al- and Ti-based thin films, retained vacancies significantly influence oxidation resistance, residual stress, and mechanical hardness \cite{ohkubo2003formation,kretschmer2021improving,euchner2015solid}.
\subsection*{Deformation and Thermomechanical Processing}

Plastic deformation provides a versatile pathway for introducing non-equilibrium vacancies in metallic systems. During straining, jogged dislocations generate vacancies, while dislocation climb and annihilation, alongside grain boundaries, act as competing sinks \cite{friedbl1975dislocation}.

The resulting vacancy supersaturation reflects a balance between production and annihilation, governed by stress, strain rate, temperature, and microstructural features such as dislocation density \cite{militzer1994modelling,robson2020deformation}. At low temperatures or high strain rates, annihilation is suppressed, allowing vacancy retention that enhances diffusion and alters phase stability \cite{sun2019precipitation,robson2020deformation,ma2019dynamic}. Although high temperatures inherently increase the equilibrium vacancy concentration, the higher diffusivity at these conditions promotes annihilation at sinks, reducing the retention of excess vacancies unless stabilization mechanisms intervene \cite{robson2020deformation}.

Stabilization through solute-vacancy binding has long been recognized as a powerful tool in alloy design. Classic studies in Al-Cu systems showed that solute additions such as Sn and In trap excess vacancies introduced during quenching, promoting the nucleation of $\theta'$-\ce{Al2Cu} during artificial aging and improving precipitation homogeneity and strength \cite{le1978solute,ringer1995effect,silcock1959effect}. Building on these insights, recent work has extended solute-vacancy stabilization to complex and nanostructured alloys. In nanograined Al-Cu-Sc systems, for example, cryogenic deformation promotes solute-vacancy complex formation, suppressing intergranular $\theta$-\ce{Al2Cu} precipitation and enhancing ductility \cite{wu2022freezing}.

Severe plastic deformation (SPD) techniques such as high-pressure torsion, cyclic loading, and equal-channel angular pressing (ECAP) can produce vacancy concentrations exceeding those generated by quenching \cite{wu2022freezing,sun2019precipitation,ma2019dynamic,yi2023interplay,cizek2019development}. In ultrafine-grained SPD materials, positron annihilation studies reveal vacancy clustering that influences diffusion, recovery, and precipitation behavior across systems, including Cu, Fe, and Al \cite{cizek2019development}. These effects are not transient, as solute-vacancy interactions can be deliberately tuned to guide phase transformations and precipitation pathways \cite{yi2023interplay,wu2022freezing}. For instance, in Al-Cu-In-Sb alloys, the binding strength of solutes to vacancies dictates the sequence and kinetics of $\theta'$ precipitation during aging \cite{zhang2017vacancy}.


\subsection*{Other Routes for Vacancy Generation and Stabilization}

Beyond irradiation-, shock-, deformation- and quenching-based processes, vacancies can also be generated and stabilized through less conventional but scientifically significant mechanisms. These routes, although secondary in alloy manufacturing, provide valuable insights into nonequilibrium defect formation and control.

Rapid phase transformations involving migrating interfaces, such as recrystallization, grain growth, precipitation, and martensitic transformations, can act as both sources and sinks of non-equilibrium vacancies. These arise from atomic flux imbalances and imperfect lattice reconstruction across moving boundaries, particularly when interface velocities are high \cite{upmanyu1998vacancy,soisson2022atomistic}. Both atomistic simulations and continuum models demonstrate that interface motion generates transient vacancy populations that, in turn, influence transformation pathways. Molecular dynamics studies have revealed that curvature-driven grain boundary migration emits vacancies as boundary area contracts, with vacancy clusters modulating interfacial curvature and mobility \cite{upmanyu1998vacancy}. Complementary continuum formulations predict that migrating boundaries either generate or absorb vacancies depending on the local stress state, temperature, and interface velocity \cite{mcfadden2020vacancy}. Although most of the evidence remains computational, experimental work by Chee \textit{et al.} \cite{chee2019interface} on metallic nanoparticles has directly demonstrated interface-mediated vacancy formation during interdiffusion. 

While generating vacancies is essential, their stabilization and retention are equally critical for achieving controlled microstructural evolution. The thermal history, deformation mode, and alloy chemistry collectively determine vacancy lifetimes and spatial distributions. For instance, although quenching or severe plastic deformation (SPD) may introduce excess vacancies, their persistence strongly depends on post-processing conditions and the availability of sinks such as grain boundaries and dislocations \cite{sun2019precipitation,yi2023interplay,munizaga2024thermodynamic}. Alloying plays a pivotal role in vacancy stabilization. In aluminum alloys, solute additions such as Sn and In have long been used to trap vacancies retained after quenching, thereby increasing the effective vacancy concentration available for precipitation during aging \cite{silcock1959effect,chen2023investigation,le1978solute,ringer1995effect}. This concept, established over five decades ago, has been exploited to promote uniform nucleation and enhance the distribution of strengthening phases such as $\theta'$-\ce{Al2Cu}. Similarly, in magnesium alloys, solutes such as Al and Zn stabilize vacancies following quenching, facilitating early-stage clustering and age-hardening kinetics \cite{yi2023interplay}. In contrast, solutes acting as efficient recombination centers can accelerate vacancy annihilation, reducing lifetimes and suppressing clustering \cite{daniels2021radiation}.

Emerging concepts, such as vacancy loops, offer new opportunities for strengthening without the need for second-phase particles \cite{yang2024dislocation,christensen2020vacancy}. Analogous to irradiation-induced loops, excess vacancies generated during processing may reorganize into stable nanoscale loops that hinder dislocation motion \cite{sun2019precipitation}. These structures provide a promising route to achieve high strength in single-phase alloys while mitigating corrosion challenges commonly associated with precipitates. Controlled thermal treatments, such as staged annealing after rapid quenching or deformation, may stabilize vacancy-loop architectures, marking an exciting frontier in defect-driven alloy design.

\section*{Theoretical Foundations}
\begin{figure*}[b!]
    \centering
    \includegraphics[width=.98\linewidth]{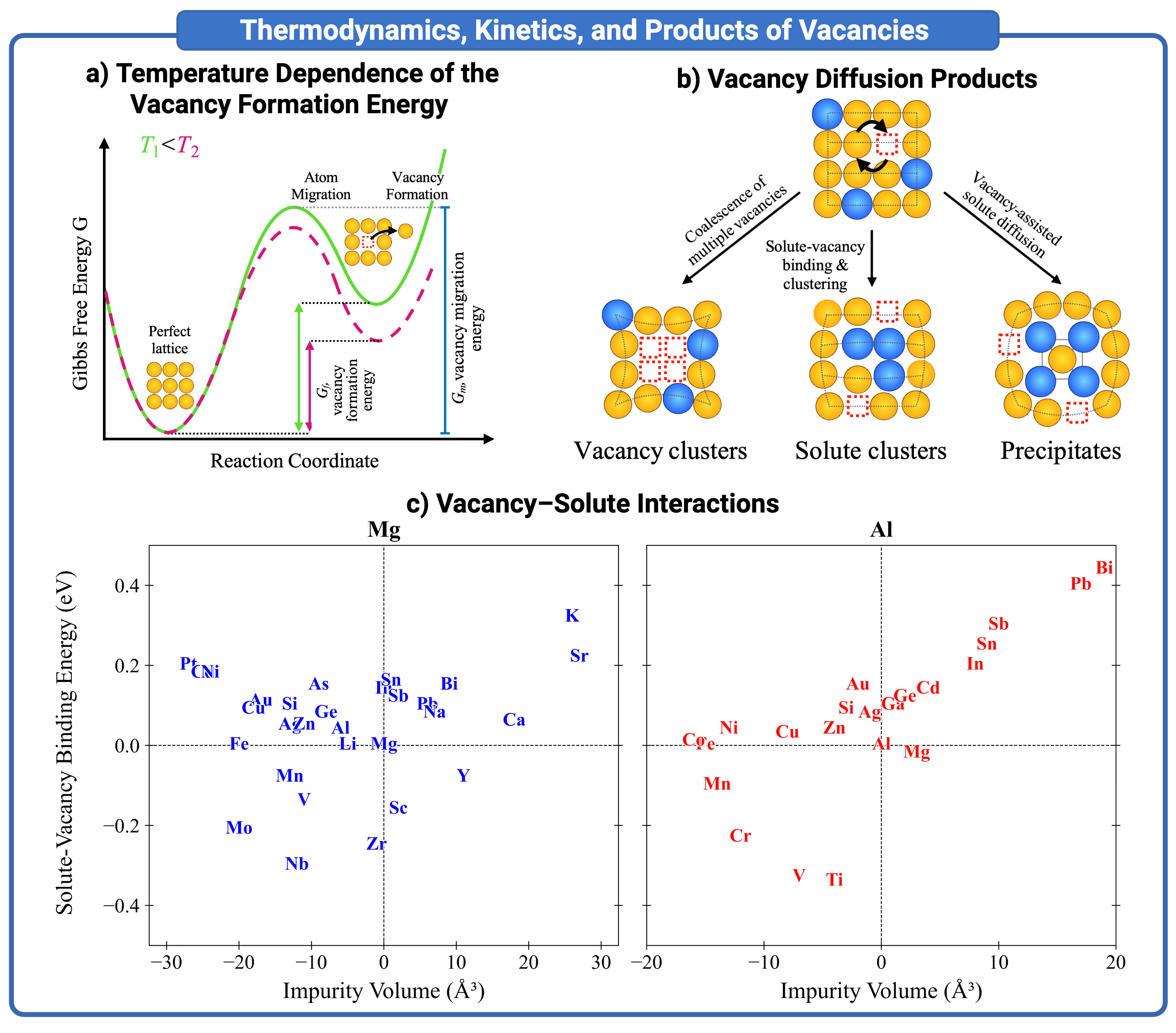}
    \caption{\justifying \textbf{Thermodynamic and kinetic principles governing vacancies in metals and alloys.} 
    \textbf{(a)} Vacancy formation is thermally activated, requiring a Gibbs free energy barrier \( G_f \), and typically occurs via atomic motion. Gibbs free energy \( G = H - T S \) decreases with temperature, promoting vacancy stabilization at high \( T \). 
    \textbf{(b)} Vacancies drive diffusion-based transformations, forming solute clusters, vacancy clusters, and precipitates through solute redistribution \cite{ma2019dynamic,peng2020solute,sun2019precipitation}. 
    Other products like stacking fault tetrahedra (SFTs), which form from partials and vacancies under irradiation, are mechanically harmful outcomes of vacancy supersaturation \cite{uberuaga2007direct}. 
    \textbf{(c)} Solute--vacancy binding energies \( E_b \) in Mg (blue) and Al (red), plotted against the change in volume produced upon addition of a single impurity into the system\cite{wolverton_solutevacancy_2007, shin2010first}. A size-dependent trend is shown: larger solutes exhibit more negative \( E_b \), indicating stronger elastic interactions. These drive clustering, segregation, and defect stabilization, which are essential to vacancy-guided alloy design.}
    \label{fig:Thermodynamics}
\end{figure*}

Equilibrium vacancy concentrations in metals are governed by thermodynamic principles. Vacancy populations may, however, be driven out of equilibrium, reflecting a balance between formation energy, entropy, and the kinetic mechanisms controlling vacancy transport, all influenced by temperature, bonding, local structure, and external driving forces. \textbf{\cref{fig:Thermodynamics}} illustrates the core thermodynamic and kinetic concepts. It summarizes vacancy formation as a thermally activated process, the increase in configurational entropy with vacancy concentration, the role of vacancy diffusion in generating clusters and precipitates, and solute-vacancy binding trends across different alloying elements in Mg and Al alloys.

This section outlines the key theoretical concepts, formation enthalpy, entropy, and local chemical effects, that determine vacancy stability and behavior in metallic systems.
\subsection*{Vacancy Thermodynamics}

The equilibrium concentration of vacancies ($C_v$) in crystalline metals is governed by a balance between the energetic cost of forming a vacancy and the entropic gain associated with transitions to thermodynamic states with increased multiplicity. This relationship is expressed as:
\begin{equation}
C_v = \exp\left(-\frac{G_f}{k_B T}\right), \quad \text{with} \quad G_f = E_f - T S_f,
\end{equation}
where $G_f$ is the Gibbs Free Energy, \(E_f\) is the enthalpy and \(S_f\) the entropy of vacancy formation. While this expression is often treated as canonical, both \(E_f\) and \(S_f\) depend sensitively on the atomic bonding environment, vibrational spectrum, and both local chemical and structural order \cite{lipnitskii2025new}.

The vacancy formation enthalpy ($E_f$) is strongly influenced by crystal structure and bonding. 
FCC metals such as Al and Ni, with dense atomic packing and more delocalized metallic bonding, 
exhibit relatively low $E_f$ values (0.7--1.4\,eV), which contribute to higher equilibrium vacancy concentrations. 
By contrast, BCC and HCP metals such as Fe, Mo, and Zr typically exhibit higher $E_f$ values (1.6--2.0\,eV). 
This trend reflects the more directional nature of bonding and reduced local bond redundancy in these lattices, 
which makes vacancy creation more energetically costly. While $E_f$ controls the equilibrium population of vacancies, diffusion kinetics depend separately on the migration enthalpy; both terms together determine self-diffusion rates in different crystal structures 
\cite{mclellan1995thermodynamics,linton2025mechanistic,alonso1989vacancy}.

The entropic contribution \(S_f\) comprises both configurational and vibrational components. While configurational entropy arises from the increase in possible atomic arrangements, vibrational entropy often dominates at elevated temperatures. Vacancy formation alters local phonon modes, typically softening low-frequency vibrations, thereby lowering the vibrational free energy and increasing \(S_f\). These effects have been captured using quasi-harmonic and anharmonic lattice dynamics models \cite{mclellan1995thermodynamics,glensk2014breakdown,bochkarev2019anharmonic,wautelet1985possible}. As a result, \(\ln C_v\) versus \(1/T\) plots often show nonlinearity near the melting point due to temperature-dependent entropy.
A comprehensive survey by Kraftmakher~\cite{kraftmakher1998equilibrium} across over 50 metallic elements confirmed that \(E_f\) scales with melting temperature, and that \(S_f\) frequently exceeds the baseline configurational value of \(k_B\), highlighting the significance of vibrational contributions. Experimental techniques such as dilatometry, positron annihilation, and electrical resistivity support these thermodynamic interpretations \cite{kraftmakher1998equilibrium}.

In chemically complex alloys (sometimes referred to as high-entropy alloys (HEAs)), vacancy thermodynamics deviates from the behavior observed in pure metals. Local atomic environments vary stochastically, creating a distribution of vacancy formation energies. First-principles studies show that such fluctuations can lower the effective energy barrier in certain configurations, resulting in elevated net vacancy concentrations \cite{wang2017thermodynamics}. Additionally, the high configurational entropy of HEAs may further promote vacancy stabilization at elevated temperatures.

Microstructural features also influence vacancy energetics. Interfaces such as grain boundaries, dislocations, and surfaces present altered bonding environments, modifying local formation energies and potentially serving as vacancy sources and sinks during non-equilibrium processing. Thermodynamic models based on Gibbs adsorption predict vacancy segregation to interfaces, such as grain boundaries, where the local enthalpy is lower than in the bulk \cite{shvindlerman1998thermodynamics}. This segregation plays a key role in grain boundary mobility, creep behavior, and sintering kinetics.

Together, these insights establish that vacancy thermodynamics are shaped by an interplay of bonding, vibrational entropy, alloy chemistry, and microstructure. Yet, despite their foundational role, thermodynamic aspects of vacancy behavior remain less explored than kinetic phenomena such as diffusion and clustering. A deeper understanding of vacancy energetics is essential for predictive control over defect populations in structural and functional metals.

\subsection*{Kinetic Principles of Vacancies}

Vacancy kinetics determine how point defects contribute to structural evolution under non-equilibrium conditions, such as energetic processing or rapid changes in thermodynamic state. Unlike thermodynamic models that predict equilibrium populations, kinetic frameworks describe how vacancies form, migrate, cluster, and annihilate at microstructural sinks over time. These dynamics are especially important following quenching, irradiation, or severe plastic deformation, where supersaturated vacancies trigger complex time-dependent processes~\cite{cizek2019development,gibbs1968vacancy}.

At the atomic scale, vacancy migration occurs through thermally activated hops between lattice sites. The frequency of such hops follows an Arrhenius relation:
\begin{equation}
\Gamma = \nu \exp\left(-\frac{G_m}{kT}\right),
\end{equation}
where $\nu$ is the attempt frequency and $G_m$ the migration barrier~\cite{le2002kinetic,schmauder2002atomistic}. This rate underpins vacancy diffusivity, self-diffusion, substitutional solute transport, recovery, and precipitation kinetics. The diffusivity of a substitutional solute, for example, depends on the vacancy concentration ($c_v$), jump distance ($a$), attempt frequency ($\nu$), and $G_m$:
\begin{equation} \label{eqn:solute-vacancy}
D_{\text{solute}} = f\, c_v\, a^2\, \nu\, \exp\left(-\frac{G_m}{kT}\right),
\end{equation}
where $f$ is the correlation factor. High vacancy concentrations accelerate clustering and reduce incubation times for nucleation.

Following a quench, mobile vacancies readily interact to form clusters such as di- and tri-vacancies and higher-order complexes, many of which become kinetically trapped and reduce long-range diffusivity~\cite{kubvena2009analysis,wang2013defect}. The evolution of such clusters can be modeled by mean-field cluster dynamics, where the concentration $c_n(t)$ of $n$-vacancy clusters evolves as
\begin{equation} \label{eqn:cluster_dynamics}
\frac{dc_n}{dt} = G_n + \sum_{i+j=n} \beta_{ij} c_i c_j - \sum_k \kappa_{nk} c_n c_k - \alpha_n c_n,
\end{equation}
with $G_n$ is the generation rate of $n$-vacancy clusters, $\beta_{ij}$ is the rate of cluster formation and $\kappa_{nk}$ is the rate of dissociation or cluster break-up, and $\alpha_n$ the annihilation rate at sinks~\cite{wang2013defect}. These equations capture how clustering progressively transfers vacancies from fast monovacancies and small mobile clusters into larger, weakly mobile or immobile aggregates, thereby depleting the mobile defect population and driving the system toward metastable states with low effective long-range diffusivity.

Beyond migration, vacancy kinetics involve multiple thermally activated steps, each characterized by an activation energy or, in heterogeneous microstructures, a distribution of energies. The vacancy formation energy ($E_f$) governs spontaneous thermal generation (\cref{fig:Thermodynamics}(a)) and, together with $E_m$, defines the effective high-temperature activation for vacancy-mediated atomic self-diffusion ($Q \approx E_f + E_m$). Binding and dissociation energies of vacancy--vacancy and vacancy--solute complexes typically regulate clustering and dissolution, shaping the forward and backward terms in \cref{eqn:cluster_dynamics}. Annihilation at sinks such as dislocations, grain boundaries, and interfaces depends on both migration to sinks and capture energetics, reflected in the coefficients $\alpha_n$. These coupled processes establish the effective kinetic coefficients ($G_n$, $\beta_{ij}$, $\alpha_n$), underscoring that defect evolution arises from an interplay of formation, migration, clustering, dissociation, and annihilation steps-each thermally activated and each capable of shaping microstructural evolution.

As vacancy clusters grow, they alter the transport landscape. Divacancies and higher-order complexes possess distinct diffusivities and binding energies, strongly affecting solute mobility and recovery rates. In early-stage aging of quenched fcc Al, simulations reveal the spontaneous formation of pentavacancies that remain immobilized from nanoseconds to seconds, trapped in deep energy basins with limited dissociation pathways~\cite{wang2013defect}. Their persistence reflects high local fluxes, reduced diffusivities, and limited sink efficiencies rather than equilibrium predictions.

At larger scales, vacancy transport couples to microstructural sinks. Dislocations, grain boundaries, and interfaces act as efficient annihilation sites, producing spatial gradients in vacancy concentration. 
For instance, experiments measuring electrical resistivity have confirmed that sink annihilation regulates defect retention and influences macroscopic properties \cite{wang2013defect}. The kinetics of this annealing process, however, are complex. Theoretical analysis suggests that, while the process should follow first-order decay, observed results often deviate, suggesting the influence of factors such as non-uniform sink distributions or competing annealing mechanisms \cite{balluffi1968annealing}. Capturing these behaviors requires time-resolved experiments and multiscale models that link atomic-scale kinetics with mesoscale redistribution.

Together, migration, clustering, annihilation, and redistribution form a hierarchy of vacancy-mediated processes that control diffusion pathways, phase transformations, and stability under non-equilibrium conditions. Far from being passive defects, vacancies act as active agents in microstructural evolution across length and time scales.

\subsection*{Vacancy-assisted solute clustering and precipitation}

Vacancies serve as both kinetic accelerators and structural mediators of phase transformations in metallic alloys. Their presence enhances local solute mobility, lowers nucleation barriers, and enables the formation of metastable or otherwise inaccessible intermediate phases. In substitutional alloys, vacancies not only facilitate solute diffusion but also alter the thermodynamic landscape, creating favorable sites for nucleation of intermetallic particles and promoting clustering and precipitation through strong solute--vacancy interactions \cite{russell1969role,xi2025multiscale}.

The terms clusters, embryos, and nuclei are sometimes used to denote distinct stages in solute aggregation. As framed by Nie \cite{nie2002roles}, clusters are sub-critical and metastable, embryos are near-critical and semi-coherent, 
and nuclei are stable and growth-favored. We hypothesize that clarifying these distinctions is particularly important in Al and Mg alloys, where vacancy-assisted clustering may govern early-stage transformations and influence both simulation predictions and interpretation of aging behavior.

In Al alloys, the formation of Guinier--Preston (GP) zones has long been linked to quenched-in vacancies\cite{chen2023investigation}. Rapid quenching enables vacancy supersaturation; these excess vacancies may help facilitate solute clustering before coherent or semi-coherent precipitates form \cite{chen2023investigation,le1978solute,bourgeois2020transforming,wang2013defect}. Studies using positron annihilation and electrical resistivity have shown that excess vacancies generated during quenching stabilize early solute clusters and extend their persistence during aging\cite{siegel1980positron,siegel1966measurement,sun2019precipitation}. Atomistic simulations reveal that pentavacancies can serve as stable nuclei, forming bcc-like configurations within the fcc aluminum lattice\cite{wang2013defect}. These defects can also trap solutes like Zn and Al in Mg alloys, increasing the likelihood of heterogeneous nucleation\cite{yi2023interplay}.

Over longer timescales, excess vacancies influence precipitate coarsening through Ostwald ripening. Vacancies carry solute atoms from smaller to larger particles by lattice diffusion, changing size distributions and usually reducing mechanical strength during thermal exposure \cite{MUKHERJEE1998101,fazeli_role_2008,roussel2001vacancy}. This process, although thermodynamically favorable, reduces hardness and mechanical stability over time, posing challenges for applications at elevated temperatures.

The thermodynamics of solute-vacancy complexes also matter. When vacancies bind strongly to solute atoms, the total equilibrium vacancy concentration increases, enhancing cluster formation and precipitation. This effect can be quantified as:
\begin{equation}
c_{\text{eq}}^{\text{total}} = c_{\text{eq}} \left(1 + Z c_{\text{sol}} \exp\left(\frac{W}{kT}\right) \right),
\end{equation}
where $Z$ is the coordination number, $c_{\text{sol}}$ is the solute concentration, and $W$ is the binding energy between solute and vacancy~\cite{chen2023investigation}. Alloys containing solutes with high binding energy promote higher effective vacancy populations and support early-stage clustering \cite{wolverton_solutevacancy_2007,shin2010first}.

Hydrogen also interacts strongly with vacancies, forming stable hydrogen-vacancy complexes that modify local structure and diffusion behavior \cite{xing2014unified}. \textit{Ab initio} calculations reveal that a single vacancy can trap multiple hydrogen atoms, lowering system energy through cooperative binding and even stabilizing molecular hydrogen within the vacancy cavity \cite{liu2009vacancy}. These complexes reduce migration barriers, alter solute diffusivity, and modify local elastic fields, thereby influencing clustering and phase transformation kinetics. Experimental and computational studies further show that in tungsten, zirconium, and nickel-based systems, such hydrogen-vacancy interactions can stabilize excess vacancies under hydrogenating or plasma environments, coupling vacancy-assisted clustering with hydrogen embrittlement and void formation \cite{liu2009vacancy,ding2022hydrogen,liu2023direct}.

In transformation-toughened and diffusional systems alike, vacancy clusters accommodate local misfit strain and reduce interface energy during phase evolution. For example, martensitic or diffusional transformations typically induce volumetric strain at the interface, leading to stress accumulation. The presence of vacancy clusters relieves such stress, allowing for smoother interface propagation and enhanced transformation continuity. This effect has been proposed in stainless steels, where interface mobility and crack resistance are tightly coupled to defect-relaxation processes \cite{laidler1972nucleation}.

Vacancy-assisted clustering enables the formation, stabilization, and evolution of complex microstructures. Beyond controlling solute mobility, vacancies can influence the functional form of clustering kinetics. In Al-Mg-Si alloys, the kinetics deviate from diffusion-controlled power-law behavior, as shown by Lay et al. \cite{lay2012vacancy}. Their positron annihilation and resistivity measurements reveal a logarithmic time dependence consistent with models where vacancies intermittently escape clusters, retrieve solute atoms, and return, thereby mediating solute transport through repeated trapping and release cycles. This mechanism demonstrates that vacancy trapping can alter both the rate and the nature of clustering behavior. Vacancies also affect solute redistribution, stress relaxation, precipitation, and coarsening, making them valuable parameters for designing and tailoring microstructures in advanced metallic systems.

\section*{Characterization Approaches}

Understanding vacancy behavior in metals and alloys requires a multi-modal toolkit that integrates high-resolution experiments with predictive modeling across scales. A range of complementary methods has been developed to quantify vacancy formation, clustering, and interactions with solutes or defects. 

As detailed in \textbf{Box 1}, experimental techniques such as PAS, APT, FIM, XRD, Electrical Resistivity, and 4D-STEM each offer distinct insights, from direct detection of open-volume defects to spatial mapping of solute clustering and local lattice strain \cite{mills2023nanoscale,speicher1966observation,chen2023investigation,simmons1960measurement,siegel1966measurement}. These techniques also help interpret changes in bulk properties and provide benchmarks for calibrating models during quenching, deformation, or thermal processing.

In parallel, \textbf{Box 2} outlines how computational and data-driven approaches, including DFT, MD, KMC, and rare-event sampling, have advanced our understanding of vacancy energetics, diffusion, and clustering \cite{wolverton_solutevacancy_2007,munizaga2024thermodynamic}. These models reveal mechanisms that are difficult to probe experimentally and are increasingly powered by machine learning frameworks trained on first-principles data \cite{mosquera2024machine,nonaka2025machine}.

Correlative workflows that integrate these tools are now central to tracing the evolution of vacancies during processing or in service. PAS can estimate vacancy concentrations and cluster sizes, while APT captures solute clustering and segregation driven by vacancy diffusion \cite{selim2021positron,weber2014vacancies,devaraj2018three}. Combined with modeling, these methods allow spatially and temporally resolved insights that refine mechanistic understanding.

No single method captures the full spectrum of vacancy phenomena. PAS offers high sensitivity to open-volume defects, APT maps local chemistry, 4D-STEM detects strain signatures, and simulations explore the energetics and stability of vacancies \cite{devaraj2018three,siegel1966measurement,mills2023nanoscale,yi2023interplay}. A comparative summary of these techniques, including their detection principles, spatial resolution, and characteristic insights, is provided in \textbf{Supplementary Table~S1}. Moving forward, progress in vacancy engineering will rely on integrating such datasets to connect atomic-scale behavior with macroscopic alloy performance. \\[-10em]

\onecolumn
\clearpage
\begin{tcolorbox}[
  colframe=black!75!gray,
  colback=white,
  fonttitle=\bfseries,
  breakable
]
\textbf{\centering \Large Box 1: Experimental Techniques for Vacancy Detection}
\vspace{5pt}

\small{\textbf{From bulk measurements to atomic-scale imaging, complementary techniques capture how vacancies shape microstructural evolution in metals and alloys.} These techniques span multiple length scales, from bulk measurements of vacancy concentration to atomic-resolution imaging of vacancy clusters and associated strain. Together, they form a coherent framework for visualizing, quantifying, and understanding vacancy behavior across materials systems.} 

\begin{center}
    \includegraphics[width=1\linewidth]{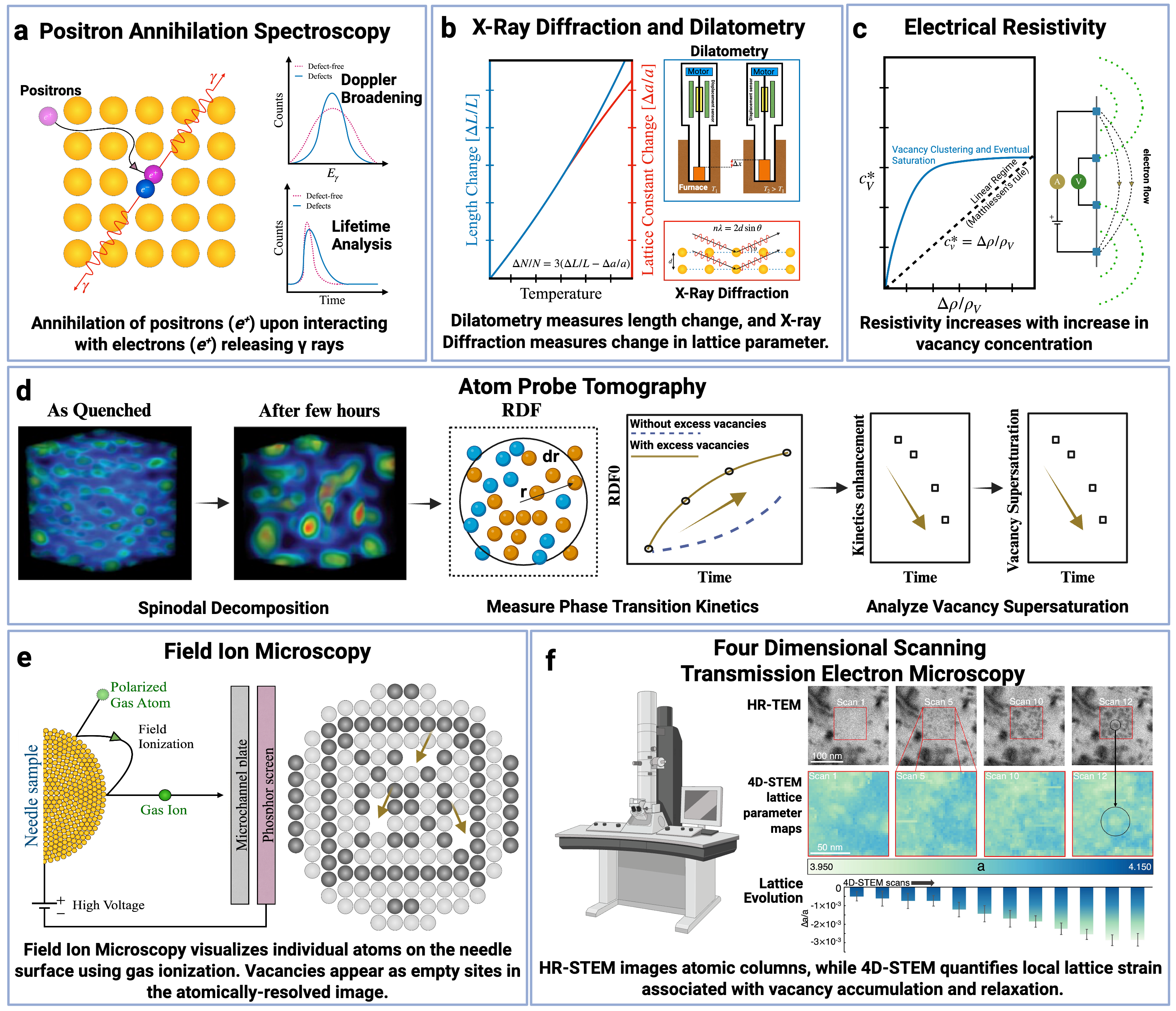}
\end{center}

\begin{flushright}
\vspace{-1em}
\scriptsize Schematics (d) and (f) adapted from \cite{chen2025using,mills2023nanoscale}.
\end{flushright}

\small
\setstretch{0.9}
\begin{multicols}{2}

\textbf{\small X-Ray Diffraction (XRD) and Dilatometry}\\
X-ray diffraction (XRD) and dilatometry have historically been used to infer vacancy concentrations from lattice expansion \cite{simmons1960measurement,simmons1962measurement}. While high-temperature lattice shifts can reflect equilibrium $C_v$, both techniques lack chemical or site specificity and require exceptional angular resolution and \textit{in situ} tracking to resolve the minute changes associated with vacancies.\\

\textbf{\small Atom Probe Tomography (APT)}\\
APT reconstructs 3D atomic maps with a spatial resolution of 0.3 nm. Vacancy effects can be inferred from solute clustering in systems with irradiated or quenched-in vacancies\cite{devaraj2018three,chen2025using,chen2023investigation}. Artifact correction is critical for accurate interpretation. Machine learning (ML) is increasingly applied for artifact correction and clustering analysis \cite{li_gault_ml_2026}.\\

\textbf{\small Electrical Resistivity}\\
Vacancies increase electron scattering, altering resistivity. Thermal cycling reveals formation, clustering, and recovery behavior, though signals must be decoupled from solute and dislocation effects\cite{colanto2010electrical,berger1973quantitative,siegel1966measurement}. Recent work employing in-situ resistivity measurements during deformation has enabled monitoring of defect evolution in metals, and such approaches could potentially be extended to study vacancy formation and migration\cite{saberi2021new}.\\ 

\columnbreak

\textbf{\small Positron Annihilation Spectroscopy (PAS)}\\
PAS estimates vacancies and binding through positron lifetime and Doppler broadening. Annihilation rates and momentum distributions reveal vacancy concentration and clustering, though interpretation requires careful modeling of positron trapping \cite{selim2021positron,dupasquier2004studies,cizek2018characterization,weber2014vacancies,cizek2019development}.\\

\textbf{\small Field Ion Microscopy (FIM)}\\
FIM provides atomic-resolution imaging by ionizing gas atoms near a field-emitter tip. Vacancy clusters appear as contrast-loss regions in cryo-prepared metals\cite{berger1973quantitative}. Recent advances in three-dimensional FIM now enable subsurface imaging and tomographic reconstruction of vacancies and lattice defects with atomic precision \cite{katnagallu2023ab,dagan2017automated}. ML-assisted reconstruction enables 3D defect visualization\cite{katnagallu2018advanced}.\\

\textbf{\small Four-Dimensional STEM (4D-STEM)}\\
4D-STEM maps nanoscale strain fields via nanobeam diffraction, revealing distortions from vacancy clusters or solute-vacancy interactions\cite{mills2023nanoscale,yang2023one}. ML-based strain mapping now improves defect detection. The method requires careful calibration and is best suited for localized studies. ML is increasingly applied to 4D‑STEM datasets to automate the high-throughput extraction of defect, strain, and orientation information\cite{oxley2021probing}.\\
\end{multicols}

\end{tcolorbox}

\begin{tcolorbox}[
  colframe=black!75!gray,
  colback=white,
  fonttitle=\bfseries,
  breakable
]
\small
\setstretch{0.9}
\textbf{\centering \Large Box 2: Computational and Machine-Learning Methods}
\begin{multicols}{2}
\textbf{\small Density Functional Theory (DFT)}\\
DFT is the gold standard for calculating vacancy formation and solute--vacancy binding energies due to its first-principles nature \cite{wolverton_solutevacancy_2007,peng2020solute,shin2010first,yi2023interplay}. However, its high computational cost limits scalability to small systems, typically one to a few hundred atoms on today's research computing resources, and sub-nanosecond time-scales when integrated with MD.\\

\textbf{\small Empirical Molecular Dynamics (MD)}\\
Empirical MD uses classical mechanics to simulate atomic motion, enabling insights into vacancy migration and defect interactions \cite{lazarev2003molecular,xu2003molecular}. Brute-force MD provides access to kinetic processes on time-scales up to microseconds, potentially tracking millions of atoms. The accuracy of these predictions are sensitive to the quality of interatomic potentials used to model the underlying bonding. MD has captured vacancy-assisted clustering during rapid cooling or deformation \cite{mendelev2007molecular,ebina2021accelerated}. Rare events like nucleation, however, require the incorporation of acceleration techniques like TIS (see below) to overcome time-scale limitations.\\

\textbf{Monte Carlo (MC) and Kinetic Monte Carlo (KMC)}\\
MC techniques sample configurational space using equilibrium probability distributions to evaluate equilibrium properties \cite{macuglia2024free}. KMC extends this approach to simulate time-dependent processes over long timescales, enabling modeling of diffusion-driven clustering and precipitation. KMC has captured vacancy-driven age-hardening in Al-Zn-Mg alloys \cite{kolesnikov2019kinetic,xi2024kinetic}. However, its accuracy depends on the precise enumeration of all available transition rates, which are often unavailable, limiting broader applicability.\\

\textbf{\small Transition Interface Sampling (TIS)}\\
TIS is a class of enhanced sampling methods for capturing rare events, such as vacancy migration or nucleation \cite{van2005elaborating}. It merges MC with either MD or KMC: transition states are generated along a proposed progress parameter, then sampled via MD or KMC to quantify transition trajectories. Though powerful for probing transformation pathways, TIS is computationally intensive and requires a suitable progress parameter. Applications to complex alloys remain limited but offer promise for understanding vacancy-driven processes in systems like Mg \cite{munizaga2024thermodynamic}.\\

\textbf{\small Machine Learning (ML)}\\
ML models, trained on DFT datasets, predict vacancy formation energies, solute--vacancy binding, migration barriers, and interatomic potentials across wide compositional spaces at low cost \cite{choudhary2023can,pentyala2022machine}. ML has been used to understand temperature-induced vacancy clustering in Tungsten \cite{zhong2025unraveling}.  As datasets grow, ML is poised to become a key tool in accelerating defect-aware alloy development.
\end{multicols}
\vspace{-1.5em}
\begin{center}
    \includegraphics[width=1\textwidth]{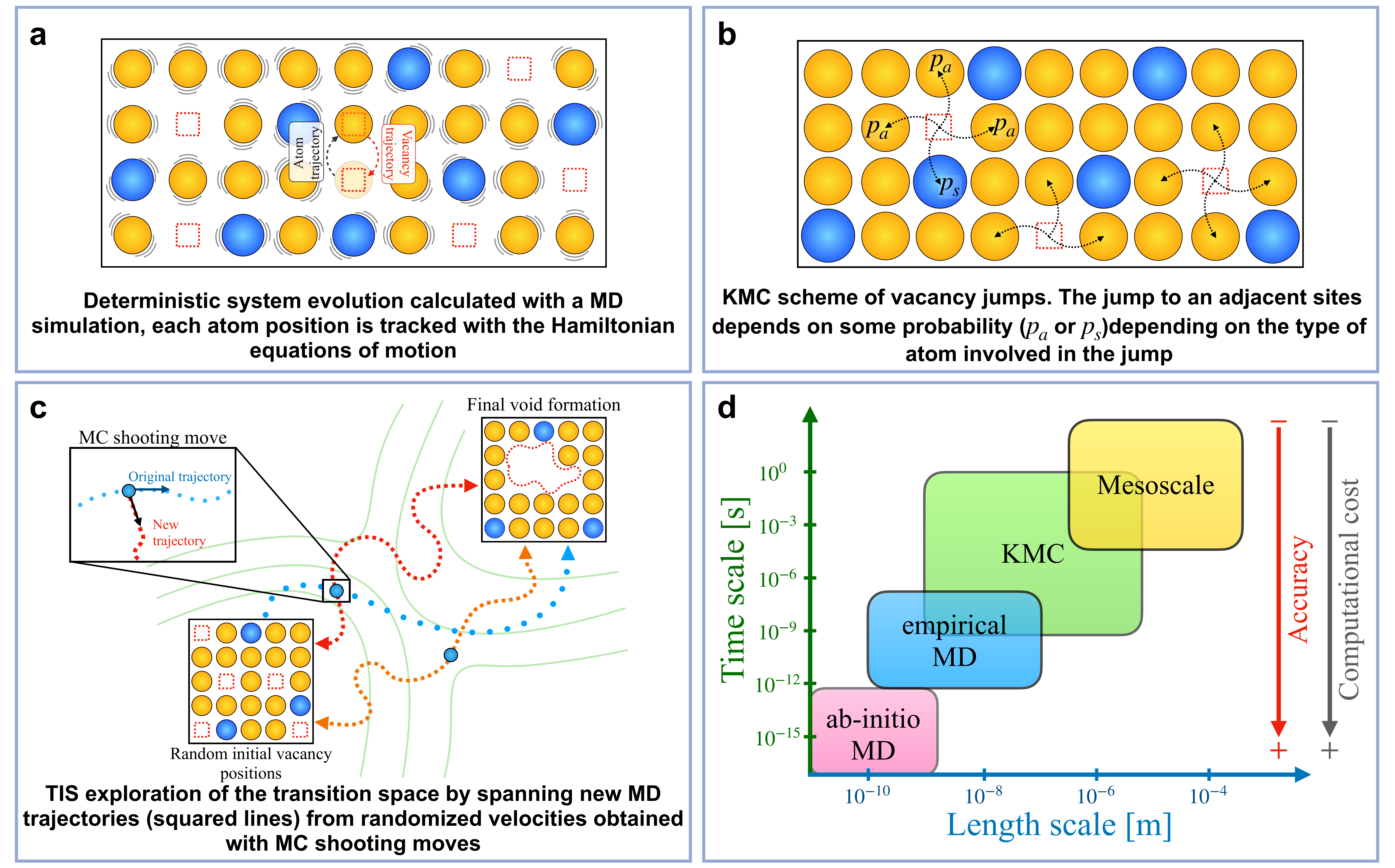}
\end{center}

\small{\textbf{Computational approaches for studying vacancy behavior across length and time scales} \textbf{(a)} Time evolution of a system calculated with MD simulations. The atomic positions at times $t_i$ are tracked using the Hamiltonian formulation of classical mechanics. \textbf{(b)} KMC scheme to find stable configurations. The method requires the reaction constants $k_i$ to compute real-time dynamics. \textbf{(c)} TIS exploration of the configurational space between two (meta)stable states. The atomic positions of two random frames from an original trajectory (blue) are used, and by randomizing their velocities, new trajectories (green and orange) are integrated using MD. A sufficient sampling will explore the entire landscape of the transition. \textbf{(d)} Computational methods comparison of length and time scales. Increased accuracy comes at a higher computational cost, which also constrains the size of the system that can be simulated.}
\end{tcolorbox}
\vspace{2em}
\twocolumn
\section*{Application Domains Enabled by Vacancy Engineering}

Building on the thermodynamic and mechanistic foundations introduced earlier, we now examine how vacancy engineering can be applied to improve metallic materials for diverse demands. Controlled manipulation of vacancy populations through thermal or deformation processing governs solute transport, phase evolution, and defect stability. These effects underpin performance gains in systems ranging from lightweight structural alloys to high-temperature components, electrical conductors, and biomedical implants. Although vacancies are also pivotal in radiation-tolerant and shock-resistant systems, this review focuses on applications outside such extreme environments.

\begin{figure*}[h!]
    \centering
    \includegraphics[width=0.7\linewidth]{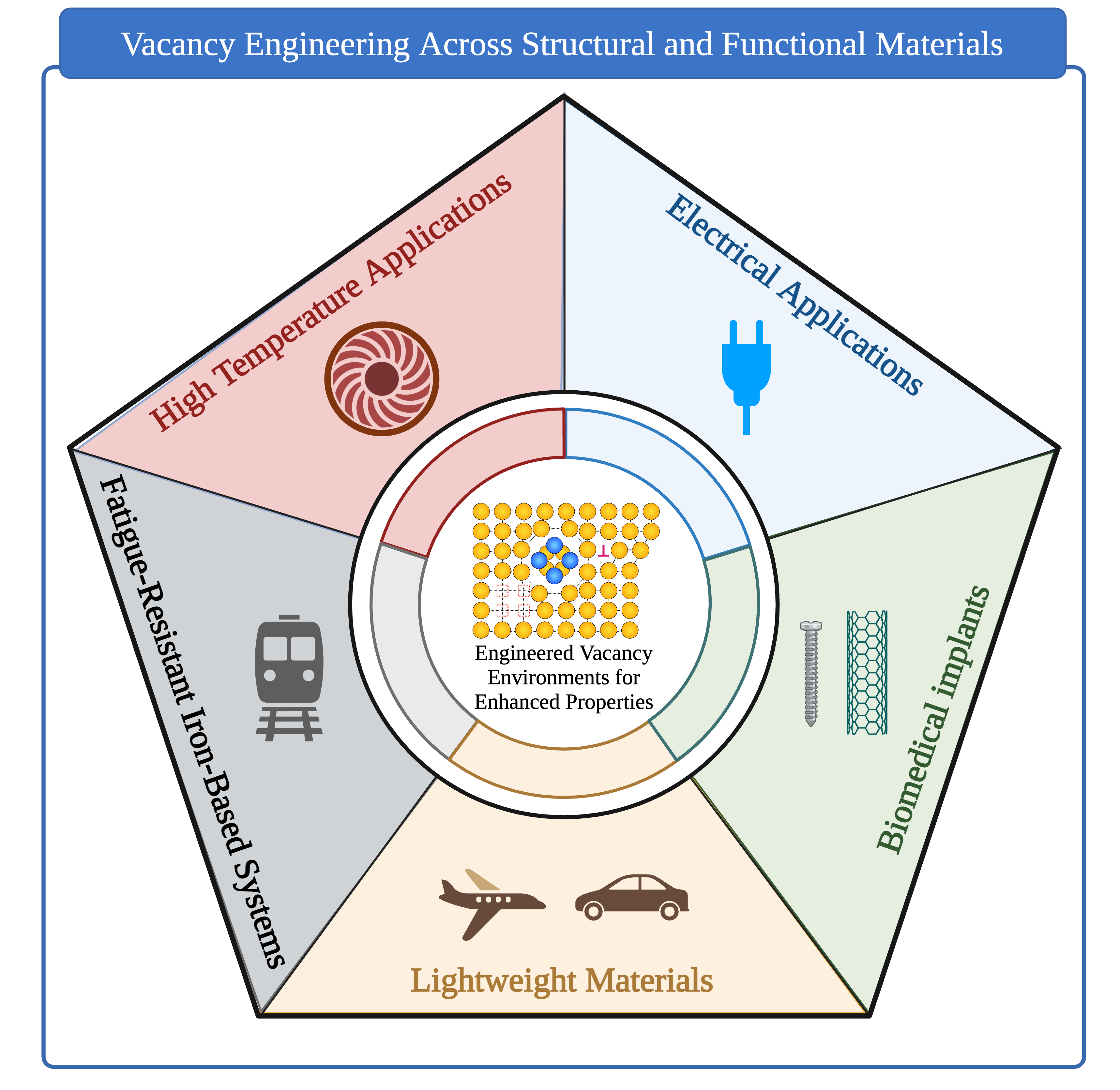}
     \caption{\justifying \textbf{Vacancy engineering enables atomic-scale control over solute partitioning, defect redistribution, and phase evolution across structural and functional material systems.}}
    \label{fig:applications_vacancies}
    \vspace{-1em}
\end{figure*}

\textbf{Figure~\ref{fig:applications_vacancies}} illustrates how vacancy-driven mechanisms influence microstructural evolution and enable application-specific property tuning. The following sections are organized by functional requirements rather than by alloy systems to highlight cross-cutting strategies.

\subsection*{Lightweight materials}
In aluminum- and magnesium-based alloys that underpin automotive and aerospace technologies, vacancies play a critical role in controlling the evolution of strengthening features that govern yield strength, fatigue resistance, and thermal stability.

In age-hardenable aluminum alloys, quenched-in vacancies promote early solute clustering and the nucleation of strengthening precipitates, thereby impeding dislocation motion and enhancing yield strength and fatigue resistance. Sustained excess-vacancy conditions lower nucleation barriers and accelerate the formation of metastable precipitates by facilitating solute redistribution \cite{bourgeois2020transforming,chen2023investigation}. As aging progresses, vacancy concentrations approach near-equilibrium values as excess vacancies are consumed through diffusion and clustering. Consequently, late-stage coarsening (overaging) is governed by solute diffusivity and interfacial energetics, consistent with diffusion-controlled Ostwald ripening models \cite{roussel2001vacancy,MUKHERJEE1998101}. Similar effects occur for deformation-generated vacancies; in alloys such as AA2024 and AA7075, cyclic loading at room temperature accelerates solute clustering and delays dislocation recovery, thereby enhancing dynamic strength \cite{sun2019precipitation}. First-principles studies by Wang et al. further indicate that compositionally tuned vacancy-solute binding energies can guide phase selection and morphology \cite{wang2024modelling}, improving mechanical response under cyclic or elevated temperature service.

Magnesium-based systems similarly benefit from non-equilibrium content of vacancies introduced through severe plastic deformation (e.g., ECAP). In Mg-Al alloys, although homogeneous clustering is thermodynamically unfavorable during conventional aging, excess vacancies can form solute-rich clusters that strengthen the material without compromising ductility \cite{yi2023interplay}.

\subsection*{High-Temperature Applications}

Vacancy engineering can be utilized to enhance the performance of high-temperature alloys by stabilizing microstructures, facilitating the formation of ordered phases, and mitigating creep and grain growth.

Tungsten alloys used in fusion and aerospace environments are prone to recrystallization despite their high melting point. Alloying with rhenium or tantalum modifies vacancy formation enthalpies, enhances solute-vacancy binding, and improves grain boundary cohesion. These interactions stabilize subgrain structures and preserve strength during thermal exposure \cite{bonny2020trends, maier1979high}.

Zirconium alloys, critical in nuclear reactors, rely on controlling vacancy-mediated diffusion to regulate second-phase precipitation and suppress the formation of embrittling intermetallics. In hydrogen-rich environments, vacancy-hydrogen interactions alter hydride-formation kinetics, thereby contributing to dimensional stability during thermal cycling \cite{xie2022strengthening}.

In titanium aluminides (e.g., $\alpha_2$-\ce{Ti3Al} and O-phase), vacancies mediate phase transformation during aging and help maintain substructural integrity. Thermomechanical treatments introduce vacancy-rich debris structures that resist recovery, enhancing creep and oxidation resistance \cite{appel2013role}.

Ordered Fe-Al intermetallics also benefit from vacancy-controlled diffusion and ordering. Antisite defects and vacancy-solute complexes influence mechanical properties and oxidation behavior \cite{haraguchi2003determination,morris1998quenching,haraguchi2005vacancy}. Tailoring defect concentrations through thermal or compositional control improves creep resistance and microstructural stability at elevated temperatures.

\subsection*{Fatigue-Resistant and High-Stability Iron-Based Systems}

Beyond high-temperature endurance, vacancy-mediated pathways are equally critical in environments dominated by cyclic loading and mechanical fatigue. In iron-based structural components, managing the early stages of damage nucleation offers a pathway to improve longevity.

Iron-based systems benefit from vacancy engineering in fatigue-prone, high-stress applications, such as rail, turbine, and reactor components \cite{ke2025role}. Positron annihilation studies reveal that vacancy-type defects form early in fatigue life and concentrate near stress risers, initiating microcracks \cite{hori2005investigation}. In Fe-Cu alloys, vacancy-assisted Cu clustering impedes dislocation motion and enhances fatigue life. However, excessive clustering can lead to embrittlement, necessitating thermal treatments to regulate cluster density \cite{thomas1963clustering}. Pre-aging and deformation strategies guide vacancy evolution to delay crack initiation and improve durability.

\subsection*{Electrical Applications}

In addition to enhancing mechanical properties, vacancy engineering plays a critical role in systems where electronic performance is equally vital. The ability to simultaneously optimize strength and electrical conductivity is particularly advantageous for high-performance conductors and functional copper-based alloys.

In such systems, vacancy-solute interactions provide a means to finely tune this balance. For instance, in Cu-Sc and Cu-Ni alloys, clustering of vacancy-solute complexes contributes to matrix strengthening but also introduces localized electron scattering, thereby decreasing conductivity \cite{zhang2015ab,dolling2022copper}. To overcome this, an aging heat treatment can be applied to coarsen the clusters, a process that clears pathways for electrons to flow and significantly improves conductivity while retaining substantial strengthening \cite{li2007study}.

Vacancies generated during heat treatment and quenching, or during plastic deformation, can further influence work hardening. Studies reveal that vacancy clusters impede dislocation climb and dynamic recovery, elevating yield strength and modifying strain hardening behavior in Cu \cite{galligan1963effect}. However, these same defects can also serve as additional scattering centers, further impacting electron mobility \cite{zhang2015ab}. Therefore, precise control over vacancy populations and their temporal evolution is essential to optimize both mechanical and electrical performance in copper-based functional materials.

\subsection*{Biomedical Implants}

Biomedical systems present particularly challenging environments, as materials must simultaneously satisfy structural, chemical, and biological requirements. Here, vacancy engineering enables precise tuning of both mechanical behavior and physiological compatibility.

For biodegradable implants, controlling corrosion alongside mechanical integrity is essential \cite{raguraman2024machine}. Excess vacancies accelerate solute diffusion and precipitate formation, facilitating microstructural refinement that enhances strength while promoting uniform degradation. In Mg-based systems, vacancy-driven formation of phases such as \ce{Mg2Ca},
\ce{Ca2Mg6Zn3}, and \ce{Mg17Al12} modulates local electrochemical behavior, influencing degradation pathways and biocompatibility \cite{ma2019dynamic,raguraman2025microstructure,raguraman_simultaneous_2025}. Techniques such as Equal Channel Angular Pressing (ECAP) introduce high vacancy densities, which in turn promote fine-scale clustering and precipitate control, crucial for achieving both load-bearing capability and controlled resorption \cite{sasaki_deformation_2022,raguraman_simultaneous_2025}.

In permanent implants, such as Ti-6Al-4V, thermomechanical processing affects vacancy generation and recovery, thereby altering the $\alpha$-$\beta$ phase distributions. Zhou et al. \cite{zhou2021microstructural} demonstrate that vacancy-assisted diffusion of Al and V governs $\beta$ coarsening and grain refinement, resulting in superior strength, ductility, and fatigue life, which are key for orthopedic medical devices.

Austenitic stainless steels are widely used in cardiovascular stents and long-term implants. Vacancy-solute interactions here influence chromium carbide precipitation during thermal exposure. Grain boundary sensitization, a key concern in corrosion, can be mitigated by limiting vacancy mobility. Modeling studies confirm that defect control during processing helps preserve passivity and corrosion resistance in physiological environments \cite{yabuuchi2011positron, mishra2021model}.
\subsection*{Caveats and Broader Impact}

While vacancy engineering is a tool to help exert control over phase transformations and properties, its effects are not universally beneficial. In magnetic and spintronic alloys, excess vacancies can disrupt long-range chemical order, weaken magnetoelastic coupling, and suppress functional strain responses. Systems such as Heusler-type ferromagnetic shape-memory alloys are particularly sensitive to local defect structures \cite{muthui2024effect,mahmoud2013effect,kumar2024role}. Similarly, vacancy-induced symmetry breaking in spintronic materials may destabilize magnetic phases and alter electronic transport \cite{mustafa2022half,van2025effects}.

Vacancies also exhibit a dual role in corrosion. In metals, oxidation generates vacancies that diffuse along grain boundaries, promoting void formation and intergranular stress corrosion cracking\cite{arioka2008dependence,yavas2020mechanical}. Within passive oxides, vacancy migration controls ionic and electronic transport that governs film growth and breakdown. The Point Defect Model describes the interplay between cation and anion vacancy fluxes, in which balanced transport sustains passivity, but excess oxygen vacancies increase electronic leakage and trigger localized failure\cite{macdonald1992point,sikora1996new}.

Despite these caveats, vacancy engineering remains a powerful lever across structural, magnetic, and biomedical metals, generating micro-- and nano--structures that enable strength, stability, conductivity, and controlled degradation.

\section*{Challenges, Open Questions, and Future Prospects}

Building on decades of foundational research on vacancies dating back to the 1940s, our ability to harness these defects as deliberate design elements in metals and alloys holds immense potential for further expansion. One of the key bottlenecks is experimental: while techniques such as PAS, APT, and advanced electron microscopy offer complementary insights, no single method provides the combination of sensitivity and spatial resolution needed to capture vacancy populations across relevant length and time scales. Multimodal in situ workflows remain underdeveloped, and the interpretation of vacancy-sensitive signals often relies on assumptions, such as constant trapping rates, that compromise quantitative fidelity. Establishing reliable reference standards and measurement protocols is critical.

Conceptual challenges further complicate progress. Under non-equilibrium conditions, including rapid quenching, additive manufacturing, cryogenic processing, and severe plastic deformation, vacancy supersaturation can be induced. Yet the stability, lifetimes, and clustering behavior of these excess vacancies remain poorly understood. In particular, how vacancies simultaneously accelerate diffusion and stabilize favorable microstructures is still an open question. This uncertainty limits our ability to predict and design processing pathways that intentionally leverage vacancy-mediated mechanisms.

Correlating vacancy behavior with macroscopic properties, such as strength, ductility, creep resistance, and corrosion, is equally complex. Vacancies interact dynamically with solutes, dislocations, grain boundaries, and precipitates, forming an evolving microstructural network. Isolating the individual contributions of vacancies within this coupled system is non-trivial and requires carefully designed experiments with clear controls. Developing such frameworks will be critical for establishing mechanistic correlations.

Nonetheless, advances in characterization hold promise. Emerging tools, such as synchrotron-based PAS, $\gamma$-ray-induced positron annihilation lifetime spectroscopy \cite{taira2025gamma}, and 4D-STEM \cite{mills2023nanoscale}, are beginning to capture vacancy formation, migration, and annihilation under thermomechanical loading. Three-dimensional field ion microscopy (3D-FIM), combined with machine learning-assisted reconstruction, now enables atomic-scale visualization and tomographic mapping of vacancy clusters and lattice defects with unprecedented precision \cite{katnagallu2018advanced,dagan2017automated}. Correlative workflows that integrate APT, field-ion microscopy, and high-resolution TEM with vacancy-sensitive probes could reveal the spatio-temporal evolution of solute-vacancy complexes, clusters, and defect networks.

On the computational front, machine learning is becoming a powerful enabler. Surrogate models trained on first-principles data, such as DCal.app \cite{luo2025dcal}, can already predict vacancy formation energies, migration barriers, and solute binding energies across multicomponent spaces. Coupled with kinetic Monte Carlo or molecular dynamics simulations, these models offer a route to bridge processing--structure--property linkages. However, the field still lacks standardized, open datasets of vacancy energetics and kinetics across alloy systems, limiting both model generalizability and benchmarking.

Moving forward, a mindset shift is needed: vacancies should be viewed not merely as facilitators of diffusion, but as dynamic variables that can be deliberately manipulated to influence phase evolution, solute clustering, and microstructural stability. Realizing this vision will require the standardization of measurement protocols, the construction of curated databases of vacancy-related properties, and the development of integrated experiment-modeling frameworks. Most importantly, connecting atomic-scale vacancy behavior to service-relevant performance metrics, such as fatigue life, corrosion resistance, or bioresorption, will enable metallic materials that are not only engineered to perform but also capable of adapting dynamically to their environment.

Despite their foundational role in physical metallurgy, vacancies remain an underexplored frontier. This Review unifies historical and emerging insights, highlights critical gaps, and calls on the community to embrace vacancy engineering as a distinct and powerful design tool for shaping the next generation of structural and functional metals and alloys.
\section*{Acknowledgements}
\noindent Sreenivas Raguraman, Homero Reyes Pulido, Michael L. Falk, and Timothy P. Weihs acknowledge support from the National Science Foundation under Grant No. DMR-2320355. We extend special thanks to Prof. Suhas Eswarappa Prameela for his valuable input. All figures were created using BioRender.com under a licensed Premium plan, which includes a full license to publish. 
\section*{Competing interests}
\noindent No competing interests.
\footnotesize
\bibliography{references}

\begin{thebibliography}{100}
\expandafter\ifx\csname url\endcsname\relax
  \def\url#1{\texttt{#1}}\fi
\expandafter\ifx\csname urlprefix\endcsname\relax\def\urlprefix{URL }\fi
\expandafter\ifx\csname href\endcsname\relax
  \def\href#1#2{#2} \def\path#1{#1}\fi

\bibitem{zhang2021advancing}
F.~Zhang, S.~He, R.~Li, L.~Lin, D.~Ren, B.~Liu, R.~Ang, Advancing thermoelectrics by vacancy engineering and band manipulation in sb-doped snte--cdte alloys, Applied Physics Letters 119~(17) (2021).

\bibitem{wang2021vacancy}
B.~Wang, J.~Liu, S.~Yao, F.~Liu, Y.~Li, J.~He, Z.~Lin, F.~Huang, C.~Liu, M.~Wang, Vacancy engineering in nanostructured semiconductors for enhancing photocatalysis, Journal of Materials Chemistry A 9~(32) (2021) 17143--17172.

\bibitem{wu2021recent}
Z.~Wu, Y.~Zhao, W.~Jin, B.~Jia, J.~Wang, T.~Ma, Recent progress of vacancy engineering for electrochemical energy conversion related applications, Advanced Functional Materials 31~(9) (2021) 2009070.

\bibitem{mao2024situ}
J.~Mao, B.~Mei, S.~Yang, J.~Zeng, F.~Sun, W.~Chen, F.~Song, Z.~Jiang, In situ oxygen-vacancy engineering for enhancing co2 reduction activity, ACS Materials Letters 6~(12) (2024) 5375--5383.

\bibitem{ma2019dynamic}
X.~Ma, S.~E. Prameela, P.~Yi, M.~Fernandez, N.~M. Krywopusk, L.~J. Kecskes, T.~Sano, M.~L. Falk, T.~P. Weihs, Dynamic precipitation and recrystallization in {Mg-9wt.\%Al} during equal-channel angular extrusion: A comparative study to conventional aging, Acta Materialia 172 (2019) 185--199.

\bibitem{raguraman_simultaneous_2025}
S.~Raguraman, A.~Kim, T.~Ayodeji, A.~J. Griebel, D.~Bershadsky, T.~Nguyen, T.~P. Weihs, \href{https://www.sciencedirect.com/science/article/pii/S0925838825016366}{Simultaneous optimization of strength and bio-corrosion resistance in biodegradable {ZX10} magnesium alloy via thermomechanical processing and annealing}, Journal of Alloys and Compounds 1024 (2025) 180078.
\newline\urlprefix\url{https://www.sciencedirect.com/science/article/pii/S0925838825016366}

\bibitem{oleksak2018role}
R.~P. Oleksak, M.~Kapoor, D.~E. Perea, G.~R. Holcomb, {\"O}.~N. Do{\u{g}}an, The role of metal vacancies during high-temperature oxidation of alloys, npj Materials Degradation 2~(1) (2018) 25.

\bibitem{luo2025determinants}
Z.~Luo, W.~Gao, Q.~Jiang, Determinants of vacancy formation and migration in high-entropy alloys, Science Advances 11~(1) (2025) eadr4697.

\bibitem{roy2022vacancy}
A.~Roy, P.~Singh, G.~Balasubramanian, D.~D. Johnson, Vacancy formation energies and migration barriers in multi-principal element alloys, Acta Materialia 226 (2022) 117611.

\bibitem{simmons1960measurement}
R.~Simmons, R.~Balluffi, Measurement of the equilibrium concentration of lattice vacancies in silver near the melting point, Physical Review 119~(2) (1960) 600.

\bibitem{simmons1962measurement}
R.~t. Simmons, R.~Balluffi, Measurement of equilibrium concentrations of lattice vacancies in gold, Physical Review 125~(3) (1962) 862.

\bibitem{selim2021positron}
F.~Selim, Positron annihilation spectroscopy of defects in nuclear and irradiated materials-a review, Materials Characterization 174 (2021) 110952.

\bibitem{weber2014vacancies}
M.~Weber, T.~Ablekim, K.~Lynn, Vacancies in {NiTi} shape memory alloys, in: Journal of Physics: Conference Series, Vol. 505, IOP Publishing, 2014, p. 012006.

\bibitem{chen2025using}
X.~Chen, F.~De~Geuser, A.~Kwiatkowski~da Silva, C.~Liu, E.~Woods, D.~Ponge, B.~Gault, D.~Raabe, Using spinodal decomposition to investigate diffusion enhancement and vacancy population, Advanced Science (2025) 2412060.

\bibitem{wang2024effect}
Y.~Wang, X.~Chen, H.~Zhao, W.~Sun, Q.~Zhang, B.~Gault, C.~Hutchinson, Effect of cluster chemistry on the strengthening of {Al} alloys, Acta Materialia 269 (2024) 119809.

\bibitem{katnagallu2018advanced}
S.~Katnagallu, B.~Gault, B.~Grabowski, J.~Neugebauer, D.~Raabe, A.~Nematollahi, Advanced data mining in field ion microscopy, Materials Characterization 146 (2018) 307--318.

\bibitem{dagan2017automated}
M.~Dagan, B.~Gault, G.~D. Smith, P.~A. Bagot, M.~P. Moody, Automated atom-by-atom three-dimensional (3d) reconstruction of field ion microscopy data, Microscopy and Microanalysis 23~(2) (2017) 255--268.

\bibitem{mills2023nanoscale}
S.~H. Mills, S.~E. Zeltmann, P.~Ercius, A.~A. Kohnert, B.~P. Uberuaga, A.~M. Minor, Nanoscale mapping of point defect concentrations with {4D-STEM}, Acta Materialia 246 (2023) 118721.

\bibitem{yang2023one}
Y.~Yang, W.~Zhou, S.~Yin, S.~Y. Wang, Q.~Yu, M.~J. Olszta, Y.-Q. Zhang, S.~E. Zeltmann, M.~Li, M.~Jin, et~al., One dimensional wormhole corrosion in metals, Nature communications 14~(1) (2023) 988.

\bibitem{wolverton_solutevacancy_2007}
C.~Wolverton, \href{https://www.sciencedirect.com/science/article/pii/S1359645407004594}{Solute-vacancy binding in aluminum}, Acta Materialia 55~(17) (2007) 5867--5872.
\newline\urlprefix\url{https://www.sciencedirect.com/science/article/pii/S1359645407004594}

\bibitem{shin2010first}
D.~Shin, C.~Wolverton, First-principles density functional calculations for {Mg} alloys: A tool to aid in alloy development, Scripta Materialia 63~(7) (2010) 680--685.

\bibitem{xu2003molecular}
Q.~Xu, T.~Yoshiie, H.~Huang, Molecular dynamics simulation of vacancy diffusion in tungsten induced by irradiation, Nuclear Instruments and Methods in Physics Research Section B: Beam Interactions with Materials and Atoms 206 (2003) 123--126.

\bibitem{huang1989pipe}
J.~Huang, M.~Meyer, V.~Pontikis, Is pipe diffusion in metals vacancy controlled? a molecular dynamics study of an edge dislocation in copper, Physical review letters 63~(6) (1989) 628.

\bibitem{battaile2008kinetic}
C.~C. Battaile, The kinetic monte carlo method: Foundation, implementation, and application, Computer Methods in Applied Mechanics and Engineering 197~(41-42) (2008) 3386--3398.

\bibitem{voter2007introduction}
A.~F. Voter, Introduction to the kinetic monte carlo method, in: Radiation effects in solids, Springer, 2007, pp. 1--23.

\bibitem{boleininger2023microstructure}
M.~Boleininger, D.~R. Mason, A.~E. Sand, S.~L. Dudarev, Microstructure of a heavily irradiated metal exposed to a spectrum of atomic recoils, Scientific Reports 13~(1) (2023) 1684.

\bibitem{pedchenko1969concerning}
K.~Pedchenko, V.~Karasev, V.~Trikula, Concerning the effect of neutron irradiation on certain thermophysical characteristics of metals, Journal of engineering physics 17~(4) (1969) 1253--1258.

\bibitem{majeed2025vacancy}
M.~Majeed, J.~Chen, J.~Jin, C.~Li, Vacancy dependent shock response of high-entropy alloy {FeNiCrCoCu}, International Journal of Mechanical Sciences (2025) 110408.

\bibitem{hillert2002trapping}
M.~Hillert, M.~Schwind, M.~Selleby, Trapping of vacancies by rapid solidification, Acta materialia 50~(12) (2002) 3285--3293.

\bibitem{haraguchi2003determination}
T.~Haraguchi, K.~Yoshimi, H.~Kato, S.~Hanada, A.~Inoue, Determination of density and vacancy concentration in rapidly solidified {FeAl} ribbons, Intermetallics 11~(7) (2003) 707--711.

\bibitem{yang2010anomaly}
Y.~Yang, H.~Huang, S.~J. Zinkle, Anomaly in dependence of radiation-induced vacancy accumulation on grain size, Journal of nuclear materials 405~(3) (2010) 261--265.

\bibitem{gavsparova2024positron}
S.~Ga{\v{s}}parov{\'a}, V.~Kr{\v{s}}jak, P.~Noga, J.~{\v{S}}olt{\'e}s, M.~Miklo{\v{s}}, Y.~Song, M.~Petriska, S.~Sojak, D.~Va{\v{n}}a, Z.~Sz{\'a}raz, et~al., Positron annihilation spectroscopy evaluation of using proton irradiation as a surrogate for early neutron radiation damage, Journal of Nuclear Materials 596 (2024) 155111.

\bibitem{cizek2019development}
J.~{\v{C}}{\'\i}{\v{z}}ek, M.~Jane{\v{c}}ek, T.~Vlas{\'a}k, B.~Smola, O.~Melikhova, I.~RK, D.~SV, The development of vacancies during severe plastic deformation, Materials transactions 60~(8) (2019) 1533--1542.

\bibitem{sun2019precipitation}
W.~Sun, Y.~Zhu, R.~Marceau, L.~Wang, Q.~Zhang, X.~Gao, C.~Hutchinson, Precipitation strengthening of aluminum alloys by room-temperature cyclic plasticity, Science 363~(6430) (2019) 972--975.

\bibitem{wu2022freezing}
S.~Wu, H.~S. Soreide, B.~Chen, J.~Bian, C.~Yang, C.~Li, P.~Zhang, P.~Cheng, J.~Zhang, Y.~Peng, et~al., Freezing solute atoms in nanograined aluminum alloys via high-density vacancies, Nature communications 13~(1) (2022) 3495.

\bibitem{chen2023investigation}
X.~Chen, J.~R. Mianroodi, C.~Liu, X.~Zhou, D.~Ponge, B.~Gault, B.~Svendsen, D.~Raabe, Investigation of vacancy trapping by solutes during quenching in aluminum alloys, Acta Materialia 254 (2023) 118969.

\bibitem{zhou1997vacancy}
X.~Zhou, R.~Johnson, H.~Wadley, Vacancy formation during vapor deposition, Acta materialia 45~(11) (1997) 4441--4452.

\bibitem{kapinos1995model}
V.~Kapinos, D.~Bacon, Model for vacancy-loop nucleation in displacement cascades, Physical Review B 52~(6) (1995) 4029.

\bibitem{dubinko2011radiation}
V.~Dubinko, A.~Guglya, S.~Donnelly, Radiation-induced formation, annealing and ordering of voids in crystals: Theory and experiment, Nuclear Instruments and Methods in Physics Research Section B: Beam Interactions with Materials and Atoms 269~(14) (2011) 1634--1639.

\bibitem{wurschum1996characterization}
R.~W{\"u}rschum, K.~Badura-Gergen, E.~K{\"u}mmerle, C.~Grupp, H.-E. Schaefer, Characterization of radiation-induced lattice vacancies in intermetallic compounds by means of positron-lifetime studies, Physical Review B 54~(2) (1996) 849.

\bibitem{tanaka2019irradiation}
M.~Tanaka, A.~Yabuuchi, A.~Kinomura, Irradiation-induced vacancy defects and its recovery behavior in {5N}-purity tungsten and {3N}-purity tantalum, in: AIP Conference Proceedings, Vol. 2182, AIP Publishing, 2019.

\bibitem{laidler1972nucleation}
J.~Laidler, B.~Mastel, Nucleation of voids in irradiated stainless steel, Nature 239~(5367) (1972) 97--98.

\bibitem{zinkle2014designing}
S.~J. Zinkle, L.~L. Snead, Designing radiation resistance in materials for fusion energy, Annual Review of Materials Research 44~(1) (2014) 241--267.

\bibitem{reina2011nanovoid}
C.~Reina, J.~Marian, M.~Ortiz, Nanovoid nucleation by vacancy aggregation and vacancy-cluster coarsening in high-purity metallic single crystals, Physical Review B--Condensed Matter and Materials Physics 84~(10) (2011) 104117.

\bibitem{jiang2021effects}
S.~Jiang, Y.~Huang, K.~Wang, X.~Li, H.~Deng, S.~Xiao, W.~Zhu, W.~Hu, Effects of vacancies on plasticity and phase transformation in single-crystal iron under shock loading, Journal of Applied Physics 130~(1) (2021).

\bibitem{ho2007energetics}
G.~Ho, M.~T. Ong, K.~J. Caspersen, E.~A. Carter, Energetics and kinetics of vacancy diffusion and aggregation in shocked aluminium via orbital-free density functional theory, Physical Chemistry Chemical Physics 9~(36) (2007) 4951--4966.

\bibitem{adibi2020evolving}
S.~Adibi, J.~W. Wilkerson, Evolving structure--property relationships in metals with nonequilibrium concentrations of vacancies, Journal of Applied Physics 127~(13) (2020).

\bibitem{xu2016ballistic}
D.~Xu, A.~Certain, H.-J. Lee~Voigt, T.~Allen, B.~D. Wirth, Ballistic effects on the copper precipitation and re-dissolution kinetics in an ion irradiated and thermally annealed {Fe--Cu} alloy, The Journal of chemical physics 145~(10) (2016).

\bibitem{holian1998plasticity}
B.~L. Holian, P.~S. Lomdahl, Plasticity induced by shock waves in nonequilibrium molecular-dynamics simulations, Science 280~(5372) (1998) 2085--2088.

\bibitem{fukai2001superabundant}
Y.~Fukai, Y.~Shizuku, Y.~Kurokawa, Superabundant vacancy formation in {Ni--H} alloys, Journal of alloys and compounds 329~(1-2) (2001) 195--201.

\bibitem{smalinskas1993study}
K.~Smalinskas, C.~Gengsheng, J.~Haworth, D.~Tappin, I.~Robertson, M.~Kirk, A study of the probability of vacancy loop formation in {Ni-Cu} and {Ag-Pd} alloys, Journal of nuclear materials 205 (1993) 483--489.

\bibitem{adams1993void}
J.~Adams, W.~Wolfer, Void formation in rapidly-solidified metals, Acta metallurgica et materialia 41~(9) (1993) 2625--2632.

\bibitem{kino1967vacancies}
T.~Kino, J.~Koehler, Vacancies and divacancies in quenched gold, Physical Review 162~(3) (1967) 632.

\bibitem{lukavc2013vacancy}
F.~Luk{\'a}{\v{c}}, J.~{\v{C}}i{\v{z}}ek, I.~Proch{\'a}zka, Y.~Jir{\'a}skov{\'a}, D.~Jani{\v{c}}kovi{\v{c}}, W.~Anwand, G.~Brauer, Vacancy-induced hardening in {Fe-Al} alloys, in: Journal of Physics: Conference Series, Vol. 443, IOP Publishing, 2013, p. 012025.

\bibitem{kimura1959quenched}
H.~Kimura, R.~Maddin, D.~Kuhlmann-Wilsdorf, Quenched-in vacancies in noble metals-- {I} theory of decay, Acta metallurgica 7~(3) (1959) 145--153.

\bibitem{morris1998quenching}
M.~Morris, D.~Morris, Quenching and ageing effects on defects and their structures in {FeAl} alloys, and the influence on hardening and softening, Scripta materialia 38~(3) (1998) 509--516.

\bibitem{silcock1959effect}
J.~Silcock, The effect of quenching on the formation of {GP} zones and $\theta$' in {Al-Cu} alloys, Philosophical Magazine 4~(46) (1959) 1187--1194.

\bibitem{khellaf2002quenching}
A.~Khellaf, A.~Seeger, R.~M. Emrick, Quenching studies of lattice vacancies in high-purity aluminium, Materials Transactions 43~(2) (2002) 186--198.

\bibitem{hirsch1958dislocation}
P.~Hirsch, J.~Silcox, R.~Smallman, K.~Westmacott, Dislocation loops in quenched aluminium, Philosophical Magazine 3~(32) (1958) 897--908.

\bibitem{zhang2020training}
Q.~Zhang, Y.~Zhu, X.~Gao, Y.~Wu, C.~Hutchinson, Training high-strength aluminum alloys to withstand fatigue, Nature communications 11~(1) (2020) 5198.

\bibitem{haraguchi2005vacancy}
T.~Haraguchi, K.~Yoshimi, M.~Yoo, H.~Kato, S.~Hanada, A.~Inoue, Vacancy clustering and relaxation behavior in rapidly solidified {B2} {FeAl} ribbons, Acta materialia 53~(13) (2005) 3751--3764.

\bibitem{zhou2019heat}
X.~Zhou, Y.~Liu, J.~Li, C.~Li, L.~Ma, D.~Fu, Heat treatment optimization of the gas atomized {Ti/B-SS} powders, Materials Research Express 6~(6) (2019) 066554.

\bibitem{shakil2021additive}
S.~Shakil, A.~Hadadzadeh, B.~S. Amirkhiz, H.~Pirgazi, M.~Mohammadi, M.~Haghshenas, Additive manufactured versus cast {AlSi10Mg} alloy: Microstructure and micromechanics, Results in Materials 10 (2021) 100178.

\bibitem{gruber2011strain}
W.~Gruber, S.~Chakravarty, C.~Baehtz, W.~Leitenberger, M.~Bruns, A.~Kobler, C.~K{\"u}bel, H.~Schmidt, Strain relaxation and vacancy creation in thin platinum films, Physical Review Letters 107~(26) (2011) 265501.

\bibitem{lorenzin2022stress}
G.~Lorenzin, L.~P. Jeurgens, C.~Cancellieri, Stress tuning in sputter-grown {Cu} and {W} films for {Cu/W} nanomultilayer design, Journal of Applied Physics 131~(22) (2022).

\bibitem{abadias2018stress}
G.~Abadias, E.~Chason, J.~Keckes, M.~Sebastiani, G.~B. Thompson, E.~Barthel, G.~L. Doll, C.~E. Murray, C.~H. Stoessel, L.~Martinu, Stress in thin films and coatings: Current status, challenges, and prospects, Journal of Vacuum Science \& Technology A 36~(2) (2018).

\bibitem{ohkubo2003formation}
H.~Ohkubo, Y.~Shimomura, I.~Mukouda, K.~Sugio, M.~Kiritani, Formation of vacancy clusters in deformed thin films of {Al--Mg} and {Al--Cu} dilute alloys, Materials Science and Engineering: A 350~(1-2) (2003) 30--36.

\bibitem{kretschmer2021improving}
A.~Kretschmer, A.~Kirnbauer, V.~Moraes, D.~Primetzhofer, K.~Yalamanchili, H.~Rudigier, P.~H. Mayrhofer, Improving phase stability, hardness, and oxidation resistance of reactively magnetron sputtered {(Al, Cr, Nb, Ta, Ti) N} thin films by {Si}-alloying, Surface and Coatings Technology 416 (2021) 127162.

\bibitem{euchner2015solid}
H.~Euchner, P.~Mayrhofer, H.~Riedl, F.~Klimashin, A.~Limbeck, P.~Polcik, S.~Kolozsvari, Solid solution hardening of vacancy stabilized {TixW1- xB2}, Acta Materialia 101 (2015) 55--61.

\bibitem{friedbl1975dislocation}
J.~Friedel, M.~Yoshida, On dislocation jogs as sources and sinks of vacancies, Philosophical Magazine 31~(1) (1975) 229--231.

\bibitem{militzer1994modelling}
M.~Militzer, W.~Sun, J.~Jonas, Modelling the effect of deformation-induced vacancies on segregation and precipitation, Acta metallurgica et materialia 42~(1) (1994) 133--141.

\bibitem{robson2020deformation}
J.~Robson, Deformation enhanced diffusion in aluminium alloys, Metallurgical and Materials Transactions A 51~(10) (2020) 5401--5413.

\bibitem{le1978solute}
A.~Le~Claire, Solute diffusion in dilute alloys, Journal of Nuclear Materials 69 (1978) 70--96.

\bibitem{ringer1995effect}
S.~Ringer, K.~Hono, T.~Sakurai, The effect of trace additions of {Sn} on precipitation in {Al-Cu} alloys: an atom probe field ion microscopy study, Metallurgical and Materials Transactions A 26~(9) (1995) 2207--2217.

\bibitem{yi2023interplay}
P.~Yi, T.~T. Sasaki, S.~E. Prameela, T.~P. Weihs, M.~L. Falk, The interplay between solute atoms and vacancy clusters in magnesium alloys, Acta Materialia 249 (2023) 118805.

\bibitem{zhang2017vacancy}
Y.~Zhang, Z.~Zhang, N.~V. Medhekar, L.~Bourgeois, Vacancy-tuned precipitation pathways in {Al-1.7 Cu-0.025 In-0.025 Sb (at.\%)} alloy, Acta materialia 141 (2017) 341--351.

\bibitem{upmanyu1998vacancy}
M.~Upmanyu, D.~J. Srolovitz, L.~Shvindlerman, G.~Gottstein, Vacancy generation during grain boundary migration, Interface Science 6 (1998) 289--300.

\bibitem{soisson2022atomistic}
F.~Soisson, M.~Nastar, Atomistic simulations of diffusive phase transformations with non-conservative point defects, MRS Communications 12~(6) (2022) 1015--1029.

\bibitem{mcfadden2020vacancy}
G.~McFadden, W.~Boettinger, Y.~Mishin, Effect of vacancy creation and annihilation on grain boundary motion, Acta Materialia 185 (2020) 66--79.

\bibitem{chee2019interface}
S.~W. Chee, Z.~M. Wong, Z.~Baraissov, S.~F. Tan, T.~L. Tan, U.~Mirsaidov, Interface-mediated kirkendall effect and nanoscale void migration in bimetallic nanoparticles during interdiffusion, Nature communications 10~(1) (2019) 2831.

\bibitem{munizaga2024thermodynamic}
V.~Munizaga, M.~L. Falk, The thermodynamic effects of solute on void nucleation in {Mg} alloys, The Journal of Chemical Physics 161~(4) (2024).

\bibitem{daniels2021radiation}
C.~Daniels, P.~Bellon, R.~Averback, Radiation-resistant binary solid solutions via vacancy trapping on solute clusters, Materialia 20 (2021) 101261.

\bibitem{yang2024dislocation}
W.~Yang, Q.~Guo, K.~Wang, P.~Lei, H.~Hou, Y.~Zhao, Dislocation loop and irradiation-induced synergistic-competitive mechanism in {Cu}-rich precipitates: a phase-field study, Scientific Reports 14~(1) (2024) 12767.

\bibitem{christensen2020vacancy}
M.~Christensen, W.~Wolf, C.~Freeman, E.~Wimmer, R.~Adamson, M.~Griffiths, E.~Mader, Vacancy loops in breakaway irradiation growth of zirconium: Insight from atomistic simulations, Journal of Nuclear Materials 529 (2020) 151946.

\bibitem{lipnitskii2025new}
A.~Lipnitskii, V.~Maksimenko, A.~Vyazmin, A.~Kartamyshev, D.~Poletaev, A new method of calculation of the thermodynamic properties of point defects in concentrated solid solutions: An application to vnbmotaw alloy, Computational Materials Science 256 (2025) 113945.

\bibitem{mclellan1995thermodynamics}
R.~McLellan, Y.~Angel, The thermodynamics of vacancy formation in fcc metals, Acta metallurgica et materialia 43~(10) (1995) 3721--3725.

\bibitem{linton2025mechanistic}
N.~Linton, D.~S. Aidhy, Mechanistic understanding of vacancy formation energies in {FCC} concentrated alloys from dft calculations, Acta Materialia 289 (2025) 120874.

\bibitem{alonso1989vacancy}
J.~Alonso, N.~March, Vacancy formation energy in close-packed metals connected with liquid thermodynamics at melting, Physics and Chemistry of Liquids 20~(4) (1989) 235--240.

\bibitem{glensk2014breakdown}
A.~Glensk, B.~Grabowski, T.~Hickel, J.~Neugebauer, Breakdown of the arrhenius law in describing vacancy formation energies: The importance of local anharmonicity revealed by ab initio thermodynamics, Physical Review X 4~(1) (2014) 011018.

\bibitem{bochkarev2019anharmonic}
A.~S. Bochkarev, A.~van Roekeghem, S.~Mossa, N.~Mingo, Anharmonic thermodynamics of vacancies using a neural network potential, Physical Review Materials 3~(9) (2019) 093803.

\bibitem{wautelet1985possible}
M.~Wautelet, A possible thermodynamical relation between the energy and entropy of formation of defects and the melting temperature, Physics Letters A 108~(2) (1985) 99--102.

\bibitem{kraftmakher1998equilibrium}
Y.~Kraftmakher, Equilibrium vacancies and thermophysical properties of metals, Physics Reports 299~(2-3) (1998) 79--188.

\bibitem{wang2017thermodynamics}
Z.~Wang, C.~Liu, P.~Dou, Thermodynamics of vacancies and clusters in high-entropy alloys, Physical Review Materials 1~(4) (2017) 043601.

\bibitem{shvindlerman1998thermodynamics}
L.~Shvindlerman, R.~Faulkner, Thermodynamics of vacancies and impurities at grain boundaries, Interface Science 6 (1998) 213--224.

\bibitem{peng2020solute}
J.~Peng, S.~Bahl, A.~Shyam, J.~A. Haynes, D.~Shin, Solute-vacancy clustering in aluminum, Acta Materialia 196 (2020) 747--758.

\bibitem{uberuaga2007direct}
B.~P. Uberuaga, R.~G. Hoagland, A.~F. Voter, S.~M. Valone, Direct transformation of vacancy voids to stacking fault tetrahedra, Physical review letters 99~(13) (2007) 135501.

\bibitem{gibbs1968vacancy}
G.~Gibbs, Vacancy generation and the kinetics of oxidation, Philosophical Magazine 18~(156) (1968) 1175--1180.

\bibitem{le2002kinetic}
Y.~Le~Bouar, F.~Soisson, Kinetic pathways from embedded-atom-method potentials: Influence of the activation barriers, Physical review B 65~(9) (2002) 094103.

\bibitem{schmauder2002atomistic}
S.~Schmauder, P.~Binkele, Atomistic computer simulation of the formation of {Cu}-precipitates in steels, Computational materials science 24~(1-2) (2002) 42--53.

\bibitem{kubvena2009analysis}
J.~Kub{\v{e}}na, A.~Kub{\v{e}}na, O.~Caha, M.~Medu{\v{n}}a, Analysis of vacancy and interstitial nucleation kinetics in {Si} wafers during rapid thermal annealing, Journal of Physics: Condensed Matter 21~(10) (2009) 105402.

\bibitem{wang2013defect}
H.~Wang, D.~Rodney, D.~Xu, R.~Yang, P.~Veyssi{\`e}re, Defect kinetics on experimental timescales using atomistic simulations, Philosophical Magazine 93~(1-3) (2013) 186--202.

\bibitem{balluffi1968annealing}
R.~Balluffi, D.~Seidman, Annealing kinetics of vacancies to dislocations, The Philosophical Magazine: A Journal of Theoretical Experimental and Applied Physics 17~(148) (1968) 843--848.

\bibitem{russell1969role}
K.~C. Russell, The role of excess vacancies in precipitation, Scripta Metallurgica 3~(5) (1969) 313--316.

\bibitem{xi2025multiscale}
Z.~Xi, L.~G. Hector~Jr, A.~Misra, L.~Qi, Multiscale modeling of vacancy-cluster interactions and solute clustering kinetics in {Al-Mg-Zn} alloys, arXiv preprint arXiv:2507.03364 (2025).

\bibitem{nie2002roles}
J.~Nie, B.~C. Muddle, H.~I. Aaronson, S.~P. Ringer, J.~P. Hirth, On the roles of clusters during intragranular nucleation in the absence of static defects, Metallurgical and Materials Transactions A 33~(6) (2002) 1649--1658.

\bibitem{bourgeois2020transforming}
L.~Bourgeois, Y.~Zhang, Z.~Zhang, Y.~Chen, N.~V. Medhekar, Transforming solid-state precipitates via excess vacancies, Nature communications 11~(1) (2020) 1248.

\bibitem{siegel1980positron}
R.~Siegel, Positron annihilation spectroscopy, Annual Review of Materials Research 10 (1980) 393--425.

\bibitem{siegel1966measurement}
R.~Siegel, A measurement of the electrical resistivity of lattice vacancies and stacking faults in gold, The Philosophical Magazine: A Journal of Theoretical Experimental and Applied Physics 13~(122) (1966) 359--366.

\bibitem{MUKHERJEE1998101}
S.~Mukherjee, B.~R. Cooper, \href{https://www.sciencedirect.com/science/article/pii/S0921509398005115}{Coarsening in the presence of vacancies}, Materials Science and Engineering: A 248~(1) (1998) 101--114.
\newline\urlprefix\url{https://www.sciencedirect.com/science/article/pii/S0921509398005115}

\bibitem{fazeli_role_2008}
F.~Fazeli, C.~Sinclair, T.~Bastow, The {Role} of {Excess} {Vacancies} on {Precipitation} {Kinetics} in an {Al}-{Mg}-{Sc} {Alloy}, Metallurgical and Materials Transactions A 39~(10) (2008) 2297--2305.

\bibitem{roussel2001vacancy}
J.-M. Roussel, P.~Bellon, Vacancy-assisted phase separation with asymmetric atomic mobility: Coarsening rates, precipitate composition, and morphology, Physical Review B 63~(18) (2001) 184114.

\bibitem{xing2014unified}
W.~Xing, X.-Q. Chen, Q.~Xie, G.~Lu, D.~Li, Y.~Li, Unified mechanism for hydrogen trapping at metal vacancies, international journal of hydrogen energy 39~(21) (2014) 11321--11327.

\bibitem{liu2009vacancy}
Y.-L. Liu, Y.~Zhang, H.-B. Zhou, G.-H. Lu, F.~Liu, G.-N. Luo, Vacancy trapping mechanism for hydrogen bubble formation in metal, Physical Review B—Condensed Matter and Materials Physics 79~(17) (2009) 172103.

\bibitem{ding2022hydrogen}
Y.~Ding, H.~Yu, M.~Lin, K.~Zhao, S.~Xiao, A.~Vinogradov, L.~Qiao, M.~Ortiz, J.~He, Z.~Zhang, Hydrogen-enhanced grain boundary vacancy stockpiling causes transgranular to intergranular fracture transition, Acta Materialia 239 (2022) 118279.

\bibitem{liu2023direct}
S.-M. Liu, S.-H. Zhang, S.~Ogata, H.-L. Yang, S.~Kano, H.~Abe, W.-Z. Han, Direct observation of vacancy-cluster-mediated hydride nucleation and the anomalous precipitation memory effect in zirconium, Small 19~(52) (2023) 2300319.

\bibitem{lay2012vacancy}
M.~D. Lay, H.~S. Zurob, C.~Hutchinson, T.~J. Bastow, A.~Hill, Vacancy behavior and solute cluster growth during natural aging of an al-mg-si alloy, Metallurgical and Materials Transactions A 43 (2012) 4507--4513.

\bibitem{speicher1966observation}
C.~Speicher, W.~Pimbley, M.~Attardo, J.~Galligan, S.~Brenner, Observation of vacancies in the field-ion microscope, Physics letters 23~(3) (1966) 194--196.

\bibitem{mosquera2024machine}
I.~Mosquera-Lois, S.~R. Kavanagh, A.~M. Ganose, A.~Walsh, Machine-learning structural reconstructions for accelerated point defect calculations, npj Computational Materials 10~(1) (2024) 121.

\bibitem{nonaka2025machine}
Y.~Nonaka, K.~Takaki, Y.~Kobayashi, J.~Haruyama, Machine learning for predicting physical parameters of atom-vacancy defects from low-frequency noise in few-atom layer {MoS$_2$}, AIP Advances 15~(4) (2025).

\bibitem{devaraj2018three}
A.~Devaraj, D.~E. Perea, J.~Liu, L.~M. Gordon, T.~J. Prosa, P.~Parikh, D.~R. Diercks, S.~Meher, R.~P. Kolli, Y.~S. Meng, et~al., Three-dimensional nanoscale characterisation of materials by atom probe tomography, International Materials Reviews 63~(2) (2018) 68--101.

\bibitem{li_gault_ml_2026}
Y.~Li, Y.~Wei, A.~Saxena, M.~Kühbach, C.~Freysoldt, B.~Gault, \href{https://www.sciencedirect.com/science/article/pii/S0079642525001392}{Machine learning enhanced atom probe tomography analysis}, Progress in Materials Science 156 (2026) 101561.
\newline\urlprefix\url{https://www.sciencedirect.com/science/article/pii/S0079642525001392}

\bibitem{colanto2010electrical}
D.~S. Colanto, Electrical resistivity measurements to assess vacancy concentration in aluminum during ultrasonic deformation and vibratory consolidation of aluminum-carbon nanotube composites, Ph.D. thesis, Northeastern University (2010).

\bibitem{berger1973quantitative}
A.~Berger, D.~Seidman, R.~Balluffi, A quantitative study of vacancy defects in quenched platinum by field ion microscopy and electrical resistivity--i. experimental results, Acta Metallurgica 21~(2) (1973) 123--135.

\bibitem{saberi2021new}
S.~Saberi, M.~Stockinger, C.~Stoeckl, B.~Buchmayr, H.~Weiss, R.~Afsharnia, K.~Hartl, A new development of four-point method to measure the electrical resistivity in situ during plastic deformation, Measurement 180 (2021) 109547.

\bibitem{dupasquier2004studies}
A.~Dupasquier, G.~K{\"o}gel, A.~Somoza, Studies of light alloys by positron annihilation techniques, Acta Materialia 52~(16) (2004) 4707--4726.

\bibitem{cizek2018characterization}
J.~{\v{C}}{\'\i}{\v{z}}ek, Characterization of lattice defects in metallic materials by positron annihilation spectroscopy: A review, Journal of Materials Science \& Technology 34~(4) (2018) 577--598.

\bibitem{katnagallu2023ab}
S.~Katnagallu, C.~Freysoldt, B.~Gault, J.~Neugebauer, Ab initio vacancy formation energies and kinetics at metal surfaces under high electric field, Physical Review B 107~(4) (2023) L041406.

\bibitem{oxley2021probing}
M.~P. Oxley, M.~Ziatdinov, O.~Dyck, A.~R. Lupini, R.~Vasudevan, S.~V. Kalinin, Probing atomic-scale symmetry breaking by rotationally invariant machine learning of multidimensional electron scattering, npj Computational Materials 7~(1) (2021) 65.

\bibitem{lazarev2003molecular}
N.~Lazarev, V.~Dubinko, Molecular dynamics simulation of defects production in the vicinity of voids, Radiation effects and defects in solids 158~(11-12) (2003) 803--810.

\bibitem{mendelev2007molecular}
M.~I. Mendelev, B.~S. Bokstein, Molecular dynamics study of vacancy migration in {Al}, Materials Letters 61~(14-15) (2007) 2911--2914.

\bibitem{ebina2021accelerated}
H.~Ebina, S.~Fukuhara, Y.~Shibuta, Accelerated molecular dynamics simulation of vacancy diffusion in substitutional alloy with collective variable-driven hyperdynamics, Computational Materials Science 196 (2021) 110577.

\bibitem{macuglia2024free}
D.~Macuglia, Free-energy calculations in condensed matter: from early challenges to the advent of umbrella sampling, Archive for History of Exact Sciences 78~(5) (2024) 479--522.

\bibitem{kolesnikov2019kinetic}
S.~V. Kolesnikov, A.~M. Saletsky, Kinetic monte carlo simulation of small vacancy clusters electromigration on clean and defective {Cu} (100) surface, The European Physical Journal B 92 (2019) 1--6.

\bibitem{xi2024kinetic}
Z.~Xi, L.~G. Hector~Jr, A.~Misra, L.~Qi, Kinetic monte carlo simulations of solute clustering during quenching and aging of {Al--Mg--Zn} alloys, Acta Materialia 269 (2024) 119795.

\bibitem{van2005elaborating}
T.~S. Van~Erp, P.~G. Bolhuis, Elaborating transition interface sampling methods, Journal of computational Physics 205~(1) (2005) 157--181.

\bibitem{choudhary2023can}
K.~Choudhary, B.~G. Sumpter, Can a deep-learning model make fast predictions of vacancy formation in diverse materials?, AIP Advances 13~(9) (2023).

\bibitem{pentyala2022machine}
P.~Pentyala, V.~Singhania, V.~K. Duggineni, P.~A. Deshpande, Machine learning-assisted dft reveals key descriptors governing the vacancy formation energy in {Pd}-substituted multicomponent ceria, Molecular Catalysis 522 (2022) 112190.

\bibitem{zhong2025unraveling}
A.~Zhong, C.~Lapointe, A.~M. Goryaeva, K.~Arakawa, M.~Ath{\`e}nes, M.-C. Marinica, Unraveling temperature-induced vacancy clustering in tungsten: From direct microscopy to atomistic insights via data-driven bayesian sampling, PRX Energy 4~(1) (2025) 013008.

\bibitem{wang2024modelling}
X.~Wang, D.~Zhao, Y.~Xu, Y.~Li, Modelling the spatial evolution of excess vacancies and its influence on age hardening behaviors in multicomponent aluminium alloys, Acta Materialia 264 (2024) 119552.

\bibitem{bonny2020trends}
G.~Bonny, M.~Konstantinovic, A.~Bakaeva, C.~Yin, N.~Castin, K.~Mergia, V.~Chatzikos, S.~Dellis, T.~Khvan, A.~Bakaev, et~al., Trends in vacancy distribution and hardness of high temperature neutron irradiated single crystal tungsten, Acta Materialia 198 (2020) 1--9.

\bibitem{maier1979high}
K.~Maier, M.~Peo, B.~Saile, H.~Schaefer, A.~Seeger, High--temperature positron annihilation and vacancy formation in refractory metals, Philosophical Magazine A 40~(5) (1979) 701--728.

\bibitem{xie2022strengthening}
R.~Xie, C.~Xu, X.~Tian, Q.~Wang, W.~Jiang, H.~Fan, Strengthening/softening effects of vacancies on twinning deformation in zirconium, Journal of Nuclear Materials 560 (2022) 153507.

\bibitem{appel2013role}
F.~Appel, D.~Herrmann, F.~D. Fischer, J.~Svoboda, E.~Kozeschnik, Role of vacancies in work hardening and fatigue of {TiAl} alloys, International Journal of Plasticity 42 (2013) 83--100.

\bibitem{ke2025role}
J.-H. Ke, A.~Kamboj, M.~Bachhav, The role of excess vacancies in stabilizing solute clusters in low-alloy steels, Scripta Materialia 267 (2025) 116797.

\bibitem{hori2005investigation}
F.~Hori, K.~Koike, R.~Oshima, Investigation of lattice defects in the early stage of fatigue in iron by positron annihilation techniques, Applied surface science 242~(3-4) (2005) 304--312.

\bibitem{thomas1963clustering}
G.~Thomas, J.~Washburn, Clustering of vacancies and hardening, Tech. rep. (1963).

\bibitem{zhang2015ab}
X.~Zhang, M.~H. Sluiter, Ab initio prediction of vacancy properties in concentrated alloys: The case of fcc {Cu-Ni}, Physical Review B 91~(17) (2015) 174107.

\bibitem{dolling2022copper}
J.~D{\"o}lling, R.~Henle, U.~Prahl, A.~Zilly, G.~Nandi, Copper-based alloys with optimized hardness and high conductivity: Research on precipitation hardening of low-alloyed binary {CuSc} alloys, Metals 12~(6) (2022) 902.

\bibitem{li2007study}
H.~Li, S.~Xie, P.~Wu, X.~Mi, Study on improvement of conductivity of {Cu-Cr-Zr} alloys, Rare Metals 26~(2) (2007) 124--130.

\bibitem{galligan1963effect}
J.~Galligan, J.~Washburn, Effect of vacancy clusters on yielding and strain hardening of copper, Philosophical Magazine 8~(93) (1963) 1455--1466.

\bibitem{raguraman2024machine}
S.~Raguraman, M.~S. Priyadarshini, T.~Nguyen, R.~McGovern, A.~Kim, A.~J. Griebel, P.~Clancy, T.~P. Weihs, Machine learning-guided accelerated discovery of structure-property correlations in lean magnesium alloys for biomedical applications, Journal of Magnesium and Alloys 12~(6) (2024) 2267--2283.

\bibitem{raguraman2025microstructure}
S.~Raguraman, M.~Connon, C.~Byrum, R.~Berlia, V.~Ivanovskaya, B.~Ulugun, S.~E. Prameela, R.~J. Guillory~II, T.~P. Weihs, Microstructure regulates early-stage corrosion behavior and systemic aluminum fate in biodegradable mg--al alloys: Integrated in-vitro and in-vivo insights, Acta Biomaterialia (2025).

\bibitem{sasaki_deformation_2022}
T.~T. Sasaki, J.~Y. Lin, P.~Yi, Z.~H. Li, S.~E. Prameela, A.~Park, E.~Lipkin, A.~Lee, M.~L. Falk, T.~P. Weihs, K.~Hono, Deformation induced solute segregation and {G}.{P}. zone formation in {Mg}-{Al} and {Mg}-{Zn} binary alloys, Scripta Materialia 220 (2022) 114924.

\bibitem{zhou2021microstructural}
X.~Zhou, D.~Xu, S.~Geng, Y.~Fan, C.~Yang, Q.~Wang, F.~Wang, Microstructural evolution and corrosion behavior of {Ti--6Al--4V} alloy fabricated by laser metal deposition for dental applications, Journal of Materials Research and Technology 14 (2021) 1459--1472.

\bibitem{yabuuchi2011positron}
A.~Yabuuchi, M.~Maekawa, A.~Kawasuso, Positron microbeam study on vacancy generation caused by stress corrosion crack propagation in austenitic stainless steels, in: Journal of Physics: Conference Series, Vol. 262, IOP Publishing, 2011, p. 012067.

\bibitem{mishra2021model}
P.~Mishra, D.~Yavas, A.~Alshehri, P.~Shrotriya, A.~Bastawros, K.~R. Hebert, Model of vacancy diffusion-assisted intergranular corrosion in low-alloy steel, Acta Materialia 220 (2021) 117348.

\bibitem{muthui2024effect}
Z.~W. Muthui, Effect of vacancies on the electronic and magnetic properties of heusler compound {Mn$_2$CoAl}, AIP Advances 14~(1) (2024).

\bibitem{mahmoud2013effect}
N.~Mahmoud, J.~Khalifeh, B.~Hamad, A.~Mousa, The effect of defects on the electronic and magnetic properties of the {Co$_2$VSn} full heusler alloy: Ab-initio calculations, Intermetallics 33 (2013) 33--37.

\bibitem{kumar2024role}
I.~Kumar, J.~Peter, G.~Shankar, P.~Pambannan, S.~Suwas, R.~K. Biswas, R.~C. Mallik, Role of {Nb} vacancies and {Sn} substitution in modulating the thermoelectric properties of {NbCoSb}, Physical Review B 110~(20) (2024) 205207.

\bibitem{mustafa2022half}
G.~M. Mustafa, M.~Hassan, N.~M. Aloufi, S.~Saba, S.~Al-Qaisi, Q.~Mahmood, H.~Albalawi, S.~Bouzgarrou, H.~Somaily, A.~Mera, Half metallic ferroamgnetism, and transport properties of vacancy ordered double perovskites {Rb$_2$(Os/Ir)X$_6$ (X=Cl,Br)} for spintronic applications, Ceramics International 48~(16) (2022) 23460--23467.

\bibitem{van2025effects}
H.~Van~Ngoc, C.~V. Ha, J.~Guerrero-Sanchez, D.~Hoat, Effects of vacancy defects and hole doping on the electronic and magnetic properties of {InTe} monolayer: A first-principles study, Materials Science in Semiconductor Processing 198 (2025) 109699.

\bibitem{arioka2008dependence}
K.~Arioka, T.~Yamada, T.~Terachi, T.~Miyamoto, Dependence of stress corrosion cracking for cold-worked stainless steel on temperature and potential, and role of diffusion of vacancies at crack tips, Corrosion 64~(9) (2008) 691--706.

\bibitem{yavas2020mechanical}
D.~Yavas, T.~Phan, L.~Xiong, K.~R. Hebert, A.~F. Bastawros, Mechanical degradation due to vacancies produced by grain boundary corrosion of steel, Acta Materialia 200 (2020) 471--480.

\bibitem{macdonald1992point}
D.~D. Macdonald, The point defect model for the passive state, Journal of the Electrochemical Society 139~(12) (1992) 3434.

\bibitem{sikora1996new}
E.~Sikora, J.~Sikora, D.~D. Macdonald, A new method for estimating the diffusivities of vacancies in passive films, Electrochimica Acta 41~(6) (1996) 783--789.

\bibitem{taira2025gamma}
Y.~Taira, Y.~Okano, T.~Hirade, Gamma-ray-induced positron annihilation lifetime spectroscopy at {UVSOR}, in: Journal of Physics: Conference Series, Vol. 3029, IOP Publishing, 2025, p. 012022.

\bibitem{luo2025dcal}
H.~Luo, W.~Liu, Y.~Ma, C.~Liang, Dcal. app: A user-friendly tool for tracer and interdiffusion coefficient in {FCC/BCC/HCP} alloys, Calphad 89 (2025) 102811.

\end{thebibliography}
\bibliographystyle{elsarticle-num} 
\clearpage
\onecolumn
\section*{Supplementary Information}

\subsection*{Comparative Summary of Experimental Techniques to measure vacancy content}
\renewcommand{\arraystretch}{1.4} 
\begin{table}[h!]
\small
\centering
\begin{tabular}{m{4.4cm} m{3.7cm} m{2cm} m{5.2cm}}
\toprule
\textbf{Technique} & \textbf{Detection Mode} & \textbf{Resolution} & \textbf{Key Insight} \\
\midrule
Positron Annihilation Spectroscopy & Direct (PALS, DBS) & Bulk & Single vacancy concentration, vacancy cluster evolution \cite{selim2021positron, dupasquier2004studies} \\
Atom Probe Tomography & Indirect (solute clustering) & Atomic & Vacancy-mediated solute clustering pathways \cite{chen2023investigation, yi2023interplay} \\
4D-STEM & Indirect & Atomic & Local strain fields from vacancy and vacancy clusters \cite{mills2023nanoscale,yang2023one} \\
Field Ion Microscopy & Direct (atomic contrast) & Atomic & Vacancy clusters and loops in irradiated or quenched metals \cite{speicher1966observation,katnagallu2018advanced} \\
X-Ray Diffraction / Dilatometry & Indirect (lattice shifts) & Bulk & Lattice parameter evolution due to vacancies \cite{simmons1960measurement,simmons1962measurement} \\
Electrical Resistivity & Indirect (scattering) & Bulk & Vacancy formation/recovery kinetics \cite{siegel1966measurement,colanto2010electrical,berger1973quantitative} \\
\bottomrule
\end{tabular}
\caption*{\textbf{Table S1.} Comparative summary of experimental techniques for vacancy characterization. Each technique is categorized by detection mode, spatial resolution, and the key insight it offers on vacancy behavior in metals and alloys.}
\end{table}

\end{document}